\documentclass{article}
\usepackage{graphicx} 
\usepackage[utf8]{inputenc}
\usepackage{lscape}
\usepackage{float}
\usepackage{mathtools}
\usepackage{longtable}
\usepackage{amsfonts}
\usepackage{amsmath}
\usepackage{mathabx}
\usepackage{graphicx}
\usepackage{bm}
\usepackage{xcolor}
\usepackage{appendix}
\usepackage{relsize}
\usepackage{adjustbox}
\usepackage{setspace}
\usepackage{tabularray}
\usepackage{xurl}
\usepackage{hyperref}
\usepackage{authblk}  
\usepackage[numbers,sort&compress]{natbib}
\usepackage{bbm}

\usepackage{relsize}
\usepackage{booktabs}  
\usepackage{threeparttable}

\usepackage[margin=1in]{geometry}
\usepackage{multirow}

\title{A causal framework for evaluating the total effect of strategies aiming to expand screening and to improve outcomes}

\author[1]{Joy Z. Nakato\thanks{Corresponding author: \texttt{jznakato@berkeley.edu}}}
\author[2]{Janice Litunya}
\author[3]{Brian Beesiga}
\author[3]{Jane Kabami}
\author[2]{James Ayieko}
\author[3,5]{Moses R. Kamya}
\author[4]{Gabriel Chamie}
\author[1]{Laura B. Balzer}

\affil[1]{Division of Biostatistics, University of California, Berkeley, California, USA}
\affil[2]{Kenya Medical Research Institute (KEMRI), Kisumu, Kenya}
\affil[3]{Infectious Disease Research Collaboration (IDRC), Kampala, Uganda}
\affil[4]{Division of HIV, Infectious Diseases \& Global Medicine, University of California, San Francisco, California, USA}
\affil[5]{Department of Medicine, Makerere University, Kampala, Uganda}

\date{}

\begin{document}
\maketitle

\renewcommand\thefootnote{}
\footnotetext{\textbf{Abbreviations:}  ART, Antiretroviral Therapy; CRTs, Cluster Randomized  Trials; PrEP, Pre-Exposure Prophylaxis; TMLE, Targeted Minimum Loss-based Estimation.}

\newpage
\begin{abstract}

For many health conditions, there are highly efficacious treatment and prevention products. Maximizing their impact requires strategies that improve the reach of health screening in order to establish who could benefit. For example, HIV prevention strategies aim to expand risk screening and to improve uptake of pre-exposure prophylaxis (PrEP) among those experiencing risk.
Often, these  strategies  induce changes at the group-level  (e.g., health clinics or communities) and are evaluated through cluster randomized trials.
This scenario creates a complex, multilevel-mediation-missing data problem for the following reasons. First, the strategy is delivered at the cluster-level, while health screening and outcomes are at the individual-level. 
Second, the strategy improves health outcomes directly and indirectly through improved health screening. 
Third, everyone has an  ``underlying" status, which is only observed among those screened. 
To formally define the total effect in such settings, we use  Counterfactual Strata Effects: causal estimands where the outcome is only relevant for a group whose membership is subject to missingness and/or impacted by the exposure of interest.
To identify and estimate the corresponding statistical estimand, we propose a novel extension of Two-Stage targeted minimum loss-based estimation (TMLE). Simulations demonstrate the practical performance of our approach as well as the limitations of existing approaches.
\\
\end{abstract}

\noindent \textbf{Key Words: counterfactual strata effects; cluster randomized trials; group randomized trials; measurement; mediation; missing data; screening; targeted minimum loss-based estimation} 



\newpage

\section{Introduction}

Health screening is the first step to initiating a new treatment or prevention product.  
For example, HIV testing is a prerequisite for initiation of antiretroviral therapy (ART) for persons with HIV and for initiation of pre-exposure prophylaxis (PrEP) for persons at risk of HIV acquisition. 
Similarly, blood pressure must be measured prior to starting antihypertensive medication, and  LDL cholesterol must be measured prior to starting statins.
Thus, status awareness is the first pillar in the Care Cascade: the series of steps from reaching persons in need through their consistent use of treatment/prevention \cite{unaids_2024_2024,unaids_2021_prevention_cascades,unaids_hiv_nodate,unaids_2025_2025}.The term ``cascade'' highlights the drop-offs occurring at each step of the process. Indeed, the coverage of health screening is rarely 100\%. As a result, the focus population of our treatment/prevention efforts is generally unknown (i.e., subject to missingness).

To improve health at a population-level, we need strategies also improve the reach of health screening.
Such a strategy could expand outreach and service delivery into the community through village health teams or peer-based approaches (e.g., \cite{shahmanesh_effect_2021,kakande_community-based_2023,hickey_community_2025}). Alternatively, a  strategy could improve screening and uptake by integrating services at the health clinic  or by offering new services at a pharmacy (e.g., \cite{{ortblad_stand-alone_2023,kabami_multi-component_2024}}). 
These strategies are often deployed at the group-level or induces changes at the group-level. Examples of groups include communities, pharmacies, clinics, and health systems.

In these scenarios, health screening is a mediator; it is an intermediate variable on the causal pathway between the exposure (the strategy) and the outcome (uptake of treatment/prevention) \cite{greenland_causal_1999,petersen_estimation_2006,didelez_direct_2006,mackinnon_mediation_2007,robins_identifiability_1992,pearl_direct_2001}. 
Health screening can also be considered to be a competing event, because the outcome can only occur if the person participates in screening \citep{austin_introduction_2016, young_causal_2020}.
Several mediated effects may be of potential interest. Examples include controlled direct effects, natural direct and indirect effects,  stochastic  direct and indirect effects, and separable direct and indirect effects 
\cite{didelez_direct_2006,rudolph_robust_2018,vanderweele_mediation_2017,young_causal_2020,stensrud_generalized_2021}.
These estimands correspond to hypothetical interventions to change the distribution of the mediator, and many are reviewed in Rudolph et al.  \cite{rudolph_causal_2019}.
However, when studying a novel health strategy, it is of primary interest to evaluate its total causal effect, which includes the direct and indirect effects, prior to investigating mediating pathways. 

Additionally, in these settings, health screening --- the mediator --- is the key measurement indicator. All persons have an ``underlying'' health status. For example, someone is either at risk of HIV acquisition or not. However, their status is only observed if they participate in screening. Furthermore, the strategy aims to improve outcomes among \emph{everyone} in the focus population of interest (e.g., all persons with HIV risk) --- regardless of whether they participate in screening. In other words, we are interested in the strategy's effect without an additional intervention to enforce screening (i.e., eliminate the competing event) \cite{young_causal_2020}. Therefore, we cannot simply evaluate the strategy's effect among persons who screened. Additionally, standard regression-based approaches to account for differences between persons screened versus missed are not directly applicable, because  they block the indirect effect
and bias estimates of the total effect towards the null. 

A new approach is needed to estimate the total effect at a population-level, while rigorously accounting for differences between those measured and those missed. 
 
In the following, we provide a novel framework to address the \emph{multilevel-mediation-missing data problem}, which commonly arises when public health strategies  aim to improve both health screening and health outcomes. First, we give an overview of  OPAL, our motivating study, which was designed to improve PrEP uptake among persons with HIV risk in Kenya and Uganda.
Next, we describe the data generating process with a hierarchical causal model \cite{pearl_causality_2009,balzer_new_2019}. We formally define the causal estimand with Counterfactual Strata Effects, occurring when  the outcome is only relevant for a group whose membership is subject to missingness and/or impacted by the exposure \cite{balzer_evaluation_2017,balzer_far_2020, balzer_two-stage_2023, nugent_blurring_2023, gupta_mechanism_2024,petersen_causal_2024}.
For identification and estimation of the corresponding statistical estimand, we use Two-Stage targeted minimum loss-based estimation (TMLE), which separates control for missing data from effect evaluation \cite{balzer_two-stage_2023, nugent_blurring_2023}.
Our simulation study evaluates the finite sample performance of our approach and compares with common alternatives, including  Single-Stage approaches. We close with a discussion of extensions to more complex settings and  areas for ongoing research.


\section{Motivating Study}

Alcohol use is a persistent risk factor for HIV acquisition  \cite{kiwanuka_population_2017,nyabuti_characteristics_2021,goma_predicting_2024}.
In sub-Saharan Africa,  attending alcohol-serving venues as a patron or an employee has been associated with increased HIV risk \cite{mbonye_alcohol_2014,velloza_hiv-risk_2017,cain_hiv_2012,Litunya_IAS_2024,beesiga_IAS_2025}.

OPAL (Outreach and Prevention at ALcohol venues) is an ongoing cluster randomized trial (CRT) whose primary objective is to evaluate the effectiveness of community-based, recruitment strategies on PrEP uptake among adults at alcohol-serving venues in rural Kenya and Uganda (NCT05862857). The units of randomization (i.e., clusters) are groups of nearby venues. At these venues, patrons and employees receive recruitment cards offering multi-disease screening (intervention) or HIV-focused screening (control) at their local health clinic. When these cards are presented at the  clinic, the participant is offered screening  as well as  PrEP initiation if interested and eligible. The primary research question is what would be the difference in PrEP uptake among persons with HIV risk if all venues received multi-disease versus HIV-focused recruitment cards. 

Figure 1 provides a simplified schematic of OPAL and illustrates the multilevel-mediation-missing data problem inherent in studies aiming to expand health screening and improve health outcomes. First, CRTs have a hierarchical data structure; individuals are nested in clusters. Second, risk screening mediates the effect on PrEP initiation. There is a direct effect of the intervention strategy on PrEP uptake as well as an indirect effect through risk screening. Third, everyone has an ``underlying'' HIV risk status, which is only observed if they participate in screening. In other words, HIV risk status is missing unless they engage in screening, 
and our interest in improving PrEP uptake among all persons with HIV risk. Furthermore, 
persons who screen likely differ meaningfully from those who do not.
We now present our novel approach for estimating total effects, while addressing  the multilevel-mediation-missing data problem.

\begin{figure}[H]
    \centering
  \includegraphics[width=0.5\textwidth]{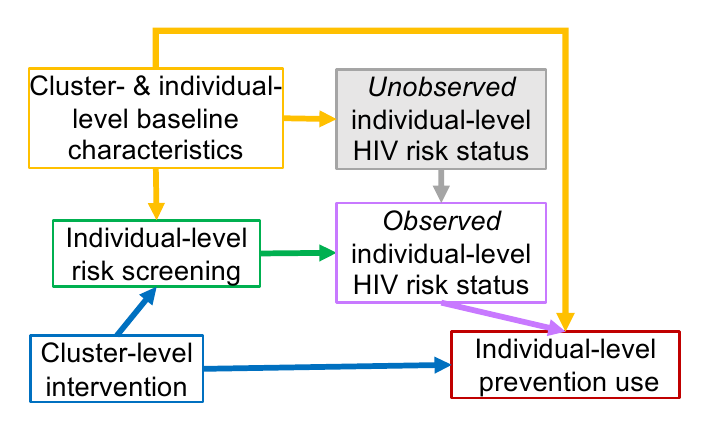}
    \caption{
Simplified causal graph for the OPAL trial to increase HIV risk screening and PrEP use among persons with HIV risk: individuals are nested within clusters; HIV risk screening mediates the
intervention effect, and  HIV risk status is missing (unobserved) if an individual does not screen. New methods are needed to
evaluate overall effectiveness, while accounting for the
multilevel-mediation-missing data problem.}
   \label{fig:opal_schematic}
\end{figure}

\section{Methods}

We first introduce the notation that will be used to formalize the data generating process, specify the causal estimand, and derive the statistical estimand. 
Throughout, clusters are indexed with $j = \{1, \ldots ,J\}$, and individuals are indexed with $i = \{1, \ldots, N_j\}$. The number of individuals in a given cluster $N_j$ often varies.  Cluster-level variables are indexed by super-script $c$ and  are shared by all individuals in that cluster. For the $j^{th}$ cluster, $E^c_j$ denotes the cluster-level baseline covariates, and $A^c_j=1$ denotes randomization to the intervention arm, while $A^c_j=0$ denotes randomization to the control arm. Bold font denotes the set of individual-level variables for a given cluster. For the $N_j$ individuals in the $j^{th}$ cluster, $\bm{W}_j$ is the matrix of individual-level characteristics at baseline: 
\begin{align*}
\bm{W}_j=
\begin{bmatrix}
W1_{1j} & W2_{1j} & \cdots  \\
W1_{2j} & W2_{2j} & \cdots  \\
\vdots & \vdots & \vdots \\
W1_{N_jj} & W2_{N_jj} & \cdots 
\end{bmatrix}
\end{align*}
In OPAL, for example, $\bm{W}_j$ could include the age, sex, alcohol use, mobility, and socioeconomic status for the $N_j$ individuals in cluster $j$. 

Asterisks denote underlying values of variables. 
Specifically, let $Y1_{ij}^*$ be in indicator that the individual $i$ in cluster $j$ is in the focus population.  In OPAL, for example, $Y1^*$ indicates being at risk of HIV acquisition.  We emphasize that $Y1^*$ is not the outcome. 
Other examples of $Y1^*$ could be indicators of having HIV for the outcome of ART initiation or  having high blood pressure for the outcome of antihypertensive medication use. 
Let  $\bm{Y1}_j^* = (Y1_{ij}^*: i = 1,\ldots ,N_j)$ be the corresponding vector of indicators for  cluster $j$.
These indicators are only observed for individuals who participate in health screening. Therefore, we define  $\bm{\Delta}_j= (\Delta_{ij}: i = 1, \ldots,N_j)$ as the vector of measurement indicators for cluster $j$. In OPAL, the HIV risk status of a given individual is observed if they are screened at the health center (i.e., if $\Delta$=1) and is missing otherwise (i.e., if $\Delta$=0). 
Then we define the observed indicators  as  $Y1_{ij}=\Delta_{ij}\times Y1_{ij}^*$  for individual $i$ in cluster $j$ and as $\bm{Y1}_j= (Y1_{ij}: i = 1, \ldots, N_j$) for cluster $j$. 
In OPAL, $Y1=1$ if the individual screened at risk for HIV acquisition and is zero otherwise (i.e., did not screen or screened but was not at risk).

Finally,  we define the outcome vector for cluster $j$ as $\bm{Y2}_j= (Y2_{ij}: i = 1, \ldots, N_j$). In OPAL, $\bm{Y2}$ consists of indicators of PrEP initiation.
Throughout, having the outcome ($Y2=1$) is only possible if the individual participates in screening and they are in the population of interest ($Y1=1$). Therefore, $Y2$ is  deterministically zero if $Y1=0$.
To match the running example where PrEP initiation occurs at the health clinic, we assume complete measurement of the outcome, but this can easily be relaxed. Altogether, we denote the observed data for  individual $i$ in cluster $j$  as $O_{ij} = (E^c_j,W_{ij}, A^c_j, \Delta_{ij},Y1_{ij},Y2_{ij} )$.

\subsection{Structural causal model}

\begin{figure}[H]
   \centering
   \includegraphics[width=0.7\textwidth]{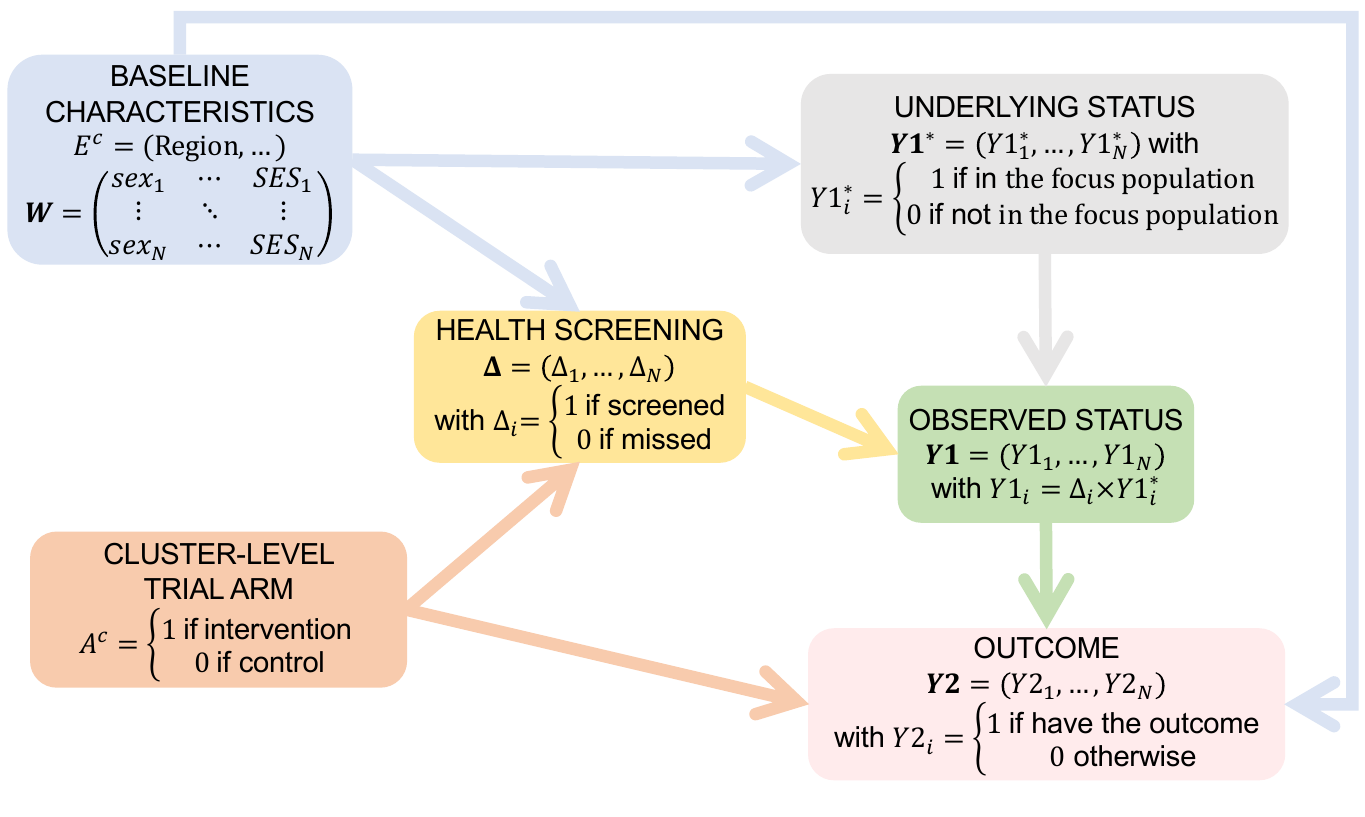}
    \caption{A simplified hierarchical causal graph to illustrate the data generating process in a CRT where the intervention strategy expands health screening and improves health outcomes. Since the causal model is defined at the cluster-level, we drop the subscript $j$ throughout. For simplicity, we have omitted the unmeasured factors: $\bm{U}=(U_{E^c}, U_{\bm{W}}, U_{A^c}, U_{\bm{Y1}^*}, U_{\bm{\Delta}},  U_{\bm{Y2}})$.}
   \label{fig:opal_dag}
\end{figure}

Causal relationships between these variables are specified through a hierarchical, non-parametric structural equation model \cite{balzer_new_2019,pearl_causality_2009}: 
 \begin{align*}
 E^c &= f_{E^c}(U_{E^c})\\
 \bm{W} &= f_{\bm{W}}(E^c, U_{\bm{W}})\\
 A^c &= f_{A^c}(U_{A^c})\\
\bm{Y1}^* &= f_{\bm{Y1}^*}(E^c, \bm{W},U_{\bm{Y1}^*})\\
\bm{\Delta} &= f_{\bm{\Delta}}(E^c,\bm{W},A^c,U_{\bm{\Delta}})\\
\bm{Y1} &= f_{Y1}(\bm{\Delta},\bm{Y1}^*) \ = \bm{\Delta} \times \bm{Y1}^* \\
\bm{Y2} &=f_{\bm{Y2}}(E^c, \bm{W},A^c,
\bm{Y1},
U_{\bm{Y2}}) 
\end{align*}
In this model, the values of the random variables on the left-hand side are determined by the structural equations 
 $(f_{E^c}, f_{\bm{W}}, f_{A^c}, f_{\bm{Y1}^*},f_{\bm{\Delta}},f_{\bm{Y1}}, f_{\bm{Y2}})$, which specify each variable's ``parents'', including  unmeasured factors $(U_{E^c},U_{\bm{W}},U_{A^c},U_{\bm{Y1}^*}, U_{\bm{\Delta}},U_{\bm{Y2}})$. Due to randomization, the unmeasured factors contributing to the intervention assignment $U_{A^c}$ are independent of the others. The corresponding directed acyclic graph is given in Figure~\ref{fig:opal_dag}.
 
We have made several exclusion restrictions. First, the trial arm $A^c$ is completely randomized; this assumption can easily be relaxed to reflect covariate-dependent randomization schemes. Second, there is no impact of the trial arm $A^c$ on underlying  status $\bm{Y1}^*$. In other words, we assume the trial arm does not change who is in the focus population (e.g., who is at risk of HIV acquisition). We relax  this and other assumptions in Section~\ref{sec:extensions}. Third, there is no effect of underlying  status $\bm{Y1}^*$ on screening $\bm{\Delta}$; this assumption is needed for identification, as detailed below (Section~\ref{sec:stage1}). Fourth, by definition, observed status  $\bm{Y1}$ is only a function of underlying status $\bm{Y1}^*$ and measurement $\bm{\Delta}$. 
Finally, the only effect of screening ($\bm{\Delta}) $ on the outcome ($\bm{Y2}$) is through establishing that the individual is in the focus population ($\bm{Y1}=1$).
Using the causal model, we can generate counterfactuals and formally define the causal estimand of interest, as  described next.

\subsection{Counterfactual Strata Effects}

Recall our goal is to estimate the total effect: a contrast in the expected counterfactual outcome under the intervention versus the expected counterfactual outcome under the control. In OPAL, for example, our goal is to evaluate the expected difference in the counterfactual probability of starting PrEP for all  persons with HIV risk if all clusters received multi-disease recruitment cards ($A^c=1$) versus HIV-only recruitment cards ($A^c=0$). Our causal effect of interest is formally defined as 
\[
\Psi^*=\mathbb{E}[Y^{c*} (1)-Y^{c*} (0)]
\]
where the expectation is over the target population of clusters and where 
\[ Y^{c*}(a^c)=\mathbb{P}[Y2(a^c)=1 \mid Y1^*=1]
\] is the counterfactual probability of starting PrEP under trial arm $a^c$, among persons at risk of HIV acquisition. Of course, we could consider alternative contrasts (e.g., ratios) and alternative summaries (e.g., weighting by cluster size) \cite{benitez_defining_2023}. Crucially, our causal parameter $\Psi^*$ is a summary of counterfactual outcomes under the two strategies of interest, but does not involve a 
hypothetical intervention to enforce measurement of status.  
In other words, $\Psi^*$ captures the direct and indirect effects of the intervention in the focus population without  enforcing 100\% compliance on screening. Therefore, $\Psi^*$ can be classified as an intent-to-treat parameter.

Due to the conditioning on underlying variables, causal parameters of this type have been termed ``Counterfactual Strata Effects'' \cite{petersen_causal_2024, gupta_mechanism_2024}.
Counterfactual Strata Effects are causal estimands where the outcome is only relevant for a group whose membership is subject to missingness and/or impacted by the exposure. In the running example, PrEP uptake is only relevant for persons who are at risk of HIV acquisition, and the population with HIV risk is generally unknown. Counterfactual Strata Effects have previously been applied to study intervention effects on population-level HIV viral suppression (the proportion of all persons with HIV who are suppressing viral replication) and the incidence of latent tuberculosis infection (the proportion of persons at risk of tuberculosis who acquire it) \cite{balzer_evaluation_2017,balzer_two-stage_2023, petersen_association_2017,havlir_hiv_2019,balzer_far_2020,nugent_blurring_2023,abbott_incident_2024,marquez_community-wide_2024}. In both examples, the outcome and focus population are subject to missingness, and the study intervention may have had an impact on focus population. (See extensions in Section~\ref{sec:extensions}.) 

Counterfactual Strata Effects are related but distinct from Principal Strata Effects \cite{grilli_nonparametric_2008-1,frangakis_principal_2002}. In the latter, effects are  defined conditional on latent classes where a post-baseline variable is used to cross-classify participants by joint potential values of that variable under each exposure level. Examples  include average causal effects among compliers, defiers, always takers, or  never takers. In OPAL, for example, we could estimate the average treatment effect among the latent subgroup of persons who would be always be screened for PrEP eligibility regardless of the recruitment card that they received. However, our goal is to evaluate the overall effectiveness of the health strategy among the entire focus population: persons with HIV risk.
Counterfactual Strata Effects are also related but distinct from the framework proposed by Young and colleagues for time-to-event outcomes with competing events \cite{young_causal_2020,stensrud_generalized_2021,martinussen_estimation_2023,janvin_causal_2024-1}.  
Throughout, health screening can be considered a competing event. Concretely, one cannot initiate PrEP without first screening for eligibility. While Young et al. study  competing events occurring over time among a well-defined cohort of participants, we center on evaluating outcomes among a focus population that is subject to missingness. In other words, our goal is to evaluate the total effect of the public health  strategy among \emph{all persons in the group of interest}, including those not reached for health screening. We now propose developing, evaluating, and applying Two-Stage TMLE for this effect.


\subsection{Two-Stage TMLE for Estimation and Inference}

Two-Stage TMLE is a general approach to identifying and estimating causal effects in a settings with missing data and dependent data \cite{balzer_two-stage_2023, nugent_blurring_2023}. In the first stage, we stratify on each cluster and focus on controlling for differential missingness/censoring of individual-level data. In the second stage, we use estimates from the first stage to evaluate the total effect and focus on improving efficiency. In the following, we propose an implementation of Two-Stage TMLE to address unique challenges when evaluating the total effect of strategies to expand health screening and to improve health outcomes. As detailed below, fully stratifying on each cluster in the first stage  simplifies the multilevel-mediation-missingness problem (Figure~\ref{fig:opal_schematic}). Specifically, we can control for differential screening without losing the effect of the cluster-level exposure (i.e., without blocking the indirect effect). 

\subsection{Stage 1: Identifying and estimating in each cluster}
\label{sec:stage1}

In Stage 1, we consider each cluster in turn. Within each cluster, the cluster-level covariates and intervention are constant. Therefore, the observed data for cluster $j$ simplify to $O_{ij}=(W_{ij}, \Delta_{ij}, Y1_{ij}, Y2_{ij})$ for $i=1,\ldots,N_j$. Equally importantly, our Stage 1 causal estimand simplifies to the counterfactual probability of the outcome for persons in the focus population: $Y^{c*}$= $\mathbb{P}(Y2=1\mid Y1^*=1)$. In OPAL, this is the  probability of initiating PrEP among persons with HIV risk. Crucially, stratifying on cluster allows the missingness mechanism to vary by cluster, while capturing the impact of the cluster-level variables, including the intervention $A^c$. 

Following \cite{balzer_evaluation_2017,balzer_far_2020, balzer_two-stage_2023, nugent_blurring_2023}, we re-express this causal estimand as the joint probability of being in the focus population and having the outcome, divided by the probability of being in the focus population:
\begin{eqnarray*}
     Y^{c*}= \frac{\mathbb{P}(Y2=1, Y1^*=1)}{\mathbb{P}( Y1^*=1)}
\label{eq:stage1_ratio}
\end{eqnarray*}
In OPAL, $Y^{c*}$ is now written as the joint probability of initiating PrEP and being at risk of HIV acquisition, divided by the prevalence of HIV risk. 

The numerator and denominator can then be identified and estimated, separately.
We first focus on the numerator. Throughout, having the outcome is contingent on being in the focus population. In OPAL, starting PrEP  is contingent upon on testing HIV-negative and reporting risk. Therefore, the numerator simplifies to the observed proportion with the outcome:
$\mathbb{P}(Y2=1)$.

This framework naturally extends to more complex settings where the outcome is also subject to missingness \cite{balzer_far_2020, balzer_two-stage_2023, nugent_blurring_2023}.

We now turn to the denominator.
Throughout, the denominator defines the focus population and is subject to  missingness: $\mathbb{P}(Y1^*=1)$.  
To identify the denominator from the observed data distribution,  we need to consider the plausibility of various missing data assumptions. Suppose the missing-completely-at-random (MCAR) held within each cluster  \cite{rubin_inference_1976}. For a given cluster in OPAL, the corresponding assumption would be that HIV risk among persons who go to the clinic for screening is perfectly representative of HIV risk among persons who do not. This is highly implausible, and we, instead, consider the missing-at-random (MAR) assumption \cite{rubin_inference_1976}: for each cluster, HIV risk among those screened is representative of risk among those missed within all possible values of the baseline covariates $W$. Importantly, the adjustment set for a given individual can include covariates of other cluster members (e.g., their friends).  Furthermore, the adjustment set for a given individual can include covariates of the sub-cluster (e.g., their household). In OPAL where a cluster is a group of nearby alcohol-serving venues, the adjustment set will include characteristics of the venue (e.g., type of alcohol served or presence of on-site rooms for sex work).
This missing data assumption  can be represented as $Y1^* \perp \Delta \mid  W$ and is further relaxed in Section~\ref{sec:extensions} to adjust for time-varying covariates, which are potentially impacted by the exposure $A^c$.
Additionally, we require that the probability of being measured within covariate values  be bounded away from 0: $\mathbb{P}( \Delta =1 \mid W=w) >0$ for all possible  $w$. Under these assumptions, the statistical estimand for the denominator is given by the G-computation formula \cite{robins_new_1986}: $\mathbb{E} [\mathbb{E}(Y1 \mid \Delta=1, W) ]$.

Altogether, our Stage 1 statistical estimand,  a cluster-level parameter, is
\begin{equation*}
Y^c = \frac{\mathbb{P}(Y2=1)}{\mathbb{E} [\mathbb{E}(Y1 \mid \Delta=1, W) ]}
\end{equation*}
In OPAL, $Y^c$ is interpreted as the probability of starting PrEP, divided the adjusted prevalence of HIV risk. To estimate the numerator, we use the empirical proportion with the outcome. To estimate the denominator, we use  TMLE, a doubly robust, substitution estimator optimizing the bias-variance trade-off for the target parameter  \cite{van_der_laan_targeted_2011}.
Within TMLE, we apply the Super Learner, an ensemble machine learning algorithm \cite{van_der_laan_super_2007}, for flexible estimation of the  the outcome regression  $\mathbb{E}(Y1 \mid \Delta=1, W)$ and measurement mechanism $\mathbb{P}(\Delta=1 \mid W)$. 
Briefly, the TMLE algorithm updates the initial estimates of $\hat{\mathbb{E}}(Y1 \mid \Delta=1, W)$ by using information from  $\hat{\mathbb{P}}(\Delta=1 \mid W)$ and obtains  targeted estimates
$\hat{\mathbb{E}}^{\star}(Y1 \mid \Delta=1, W)$, which are averaged across individuals. 
Since we stratify on each cluster in Stage 1, we need to implement the estimation algorithm $J$ times to obtain estimates for each cluster $j=\{1, \ldots,J\}$:
 \begin{eqnarray*}
 \hat{Y}_j^c =  \frac{1/N_j \sum_{i=1}^{N_j} \mathbbm{1}(Y2_{ij}=1)}{1/N_j\sum_{i=1}^{N_j}\hat{\mathbb{E}}^{\star}(Y1 \mid \Delta=1, W_{ij})}
\end{eqnarray*}

\subsection{Stage 2: Identifying and estimating the total effect}

Stage 1 focuses on identifying and estimating the expected outcome among the underlying focus population for each cluster. However, recall our primary research question is the total effect of the health strategy: 
$\Psi^*= \mathbb{E}[Y^{c*}(1)-Y^{c*}(0)]$.
In  OPAL, $\Psi^*$ is the expected difference in the counterfactual uptake of PrEP (among persons with HIV risk) if all clusters received multi-disease messaging $Y^c$(1) versus HIV-focused messaging $Y^c$(0).   
In Stage 2, we focus on evaluating the total effect $\Psi^*$ with maximum precision. In Stage 2, the observed data are at the cluster-level and given by $O^c=(E^c,W^c,A^c, \hat{Y}^c)$, where $W^c$ denote aggregates of the individual-level baseline covariates $\textbf{W}$ and $\hat{Y}^c$ are the cluster-level endpoint estimates from Stage 1.

Due to randomization of the trial arms and complete follow-up of clusters, we can obtain a point estimate of the total effect by simply contrasting the arm-specific average outcomes  (i.e., implementing an unadjusted effect estimator):

\begin{align}
\hat{\Psi}_{unadj} \  = \
\hat{\mathbb{E}}(\hat{Y}^c |A=1) - \hat{\mathbb{E}}( \hat{Y}^c | A=0) \ = \ \frac{\sum_{j=1}^J  \mathbbm{1}(A^c_j=1)\hat{Y}^c_j}{\sum_{j=1}^J \mathbbm{1}(A^c_j=1)}  - 
\frac{\sum_{j=1}^J  \mathbbm{1}(A^c_j=0)\hat{Y}^c_j}{\sum_{j=1}^J \mathbbm{1}(A^c_j=0)}
\label{eq:stage2_unadj}
\end{align}
However, statistical precision and power can be improved by adjusting for baseline covariates that are predictive of the outcome \cite{gail_design_1996,tsiatis_covariate_2008,fisher_statistical_1932,moore_covariate_2009,rosenblum_simple_2010,benitez_defining_2023,balzer_two-stage_2023}.
Therefore,  we implement a cluster-level TMLE to  obtain a more efficient estimate of the total effect.  Briefly, an initial estimate of the cluster-level outcome regression $\hat{\mathbb{E}}(\hat{Y}^c \mid A^c=a^c,E^c,W^c)$ is updated using information in the cluster-level propensity score  $\hat{\mathbb{P}}(A^c=1 \mid E^c,W^c)$. Then targeted estimates of the expected outcome under each trial arm are  averaged to obtain a point estimate:
\begin{eqnarray}
   \hat{\Psi}_{tmle}= \frac{1}{J} \sum_{j=1}^J  \hat{\mathbb{E}}^\star(\hat{Y}^c \mid A^c=1,E^c_j,W^c_j) - \frac{1}{J} \sum_{j=1}^J \hat{\mathbb{E}}^\star(\hat{Y}^c \mid A^c=0,E^c_j,W^c_j) 
\label{eq:stage2_tmle}
\end{eqnarray}

In a trial setting, we implement TMLE with Adaptive Pre-specification to data-adaptively select the adjustment approach that maximizes empirical efficiency  \cite{balzer_adaptivePre_2016,balzer_adapt_2024,balzer2025machinelearningoptimizeprecision}. In this procedure, we pre-specify candidate algorithms for estimation of the outcome regression and the known propensity score. In trials with few clusters, we recommend limiting these algorithms to ``working'' generalized linear models (GLMs) adjusting for a single covariate \cite{rosenblum_simple_2010}. In trials with many clusters, we can expand the candidate set to include more adaptive approaches adjusting for multiple covariates (e.g., stepwise regression and multivariate adaptive regression splines). The unadjusted estimator is always included as candidate. In addition, we pre-specify the loss function to be the square of the influence curve of the cluster-level TMLE; the expectation of this loss function is the variance of the TMLE \cite{rubin_empirical_2008}. Finally, we pre-specify a cross-validation scheme to evaluate performance and collaboratively select the combination of outcome regression and  propensity score estimators that minimize  the cross-validated risk estimate. (The algorithm is ``collaborative'' because the propensity score is fit in response to estimation of the outcome regression  \cite{stitelman_collaborative_2010}.) In the event that there are no gains in efficiency with any of the candidate adjustment approaches, this procedure defaults to the unadjusted estimator. 

\subsection{Statistical Inference}

Two-Stage TMLE is a multiply robust estimator. In Stage 1, TMLE offers a doubly robust estimation of $Y^c$ in each of the $J$ clusters; if either outcome regression or measurement mechanism is consistently estimated, we will have a consistent estimate of the cluster-level endpoint $Y^c$. Then in Stage 2, TMLE offers a model-robust procedure to maximize efficiency when evaluating the total effect of the cluster-level strategy.
Importantly, bias and variability in the Stage 1 $\hat{Y}^c$ estimates  contribute directly to the cluster-level influence curve and, thus, variance in Stage 2.

Under  conditions,  Two-Stage TMLE is an asymptotically linear estimator \cite{vaart_asymptotic_1998} --- 
the cluster-level TMLE minus the statistical estimand behaves in first order as an empirical mean of a mean-zero and finite variance function known as the influence curve:  $\hat{\Psi} -\Psi = \frac{1}{J}\sum^{J}_j IC_j +R_{J}$. Here, $IC_j$ is the influence curve for cluster $j$ and $R_{J}$ is a second order remainder term that goes to zero in probability. 
Then we obtain a variance estimate  by taking the sample variance of the estimated influence curve divided by the number of independent units $J$. To construct Wald-type $95\%$ confidence intervals and test the null hypothesis, we use the Student's t-distribution with $J-2$ degrees of freedom as small sample approximation to the Gaussian distribution \cite{hayes_cluster_2009}. 

The conditions for Two-Stage TMLE to be asymptotically linear are detailed in Balzer et al. \cite{balzer_two-stage_2023}. 
First, we require  that Stage 1 estimation of $Y^c$ has a negligible contribution to the remainder term $R_{J}$. 
This holds under the following conditions:  the within cluster dependence is weak enough such that the Central Limit Theorem applies in $N_j$; the ratio of the number of independent units to the smallest cluster size goes to zero (i.e., $J/min_j(N_j)\rightarrow 0$), and the individual-level estimators of the outcome regression and propensity score converge to their targets at fast enough rates such that the corresponding Stage 1 remainder term is $o_p(N_j^{-1/2})$.
Second, we require that Stage 2 estimators for the cluster-level outcome regression and measurement mechanism satisfy the usual regularity conditions \cite{balzer_two-stage_2023}.  These Stage 2 conditions are naturally satisfied when using Adaptive Pre-specification to select among working GLMs \cite{moore_covariate_2009}. While these conditions are asymptotic, our simulation study, described next, evaluates finite sample performance.

\section{Simulation Study}

To examine the finite sample performance of our proposed approach, we conducted a simulation study reflecting the multilevel-mediation-missing data problem.
In line with a systematic review revealing that most CRTs simply exclude participants with missing data \cite{fiero_statistical_2016}, we considered approaches that  restricted to participants who screened ($\Delta=1$). In line with common practice when evaluating HIV treatment/prevention strategies \cite{unaids_2024_2024,unaids_2025_2025,unaids_2021_prevention_cascades}, 
we also considered several approaches that simply restricted to persons known to be in the focus population ($Y1=1$).
For comparison to more traditional methods, we also considered two ``Single-Stage'' approaches that simultaneously account for missing data and estimate the intervention effect --- pooling individuals across clusters: generalized estimating equations (GEE) and the standard TMLE for clustered data \cite{schnitzer_effect_2014,liang_longitudinal_1986}. 
The considered estimators are described below in detail.
Simulations were conducted in \texttt{R} version 4.4.1. Code to replicate these simulations is available at 
\url{https://github.com/jznakato/multi-level-mediation-missingness}

\subsection{Generating the Data}

For each cluster $j=\{1, \ldots, J\}$, we repeated the following data generating process.  First, we generated 
cluster-level latent variables $U_{E1^c}$ and $U_{E2^c}$ by drawing independently from a $Unif(-1,1)$ and then the cluster-level observed covariates as $E1^c \sim Norm(U_{E1^c},.5)$ and $E2^c\sim Norm(U_{E2^c},.5)$. 
Next, we generated the number of individuals in the cluster as $N \sim Norm(200,10)$, rounding to the nearest whole number.  
For the $N$ individuals in the cluster, we independently generated baseline covariates: $W1\sim Norm(0,.5)$, $W2 \sim Norm(0,.5)$, and $W3\sim Bin(.5)$. We generated the underlying indicator of being in the focus population $Y1^*$ for each individual as 
$Y1^*= \mathbf{1}[U_{Y1^*}  < logit^{-1}(.5+W1+W2-W3+0.25E1^c+0.25E2^c)]$ with $U_{Y1^*}\sim Unif(0,1)$.
For each individual, we generated the counterfactual measurement indicator under  $a^c=\{0,1\}$ as
 $\Delta(a^c) = \mathbf{1}[U_\Delta < logit^{-1}(-.5+.6a^c+.5W1+.4W2-.4a^cW3+.4(1-a^c)W3+.1U_{E1^c}+.1U_{E2^c})]$ with $U_\Delta \sim Unif(0,1)$.
Likewise, for each individual, we generated the counterfactual outcome $Y2(a^c)$ under $a^c \in \{0,1\}$ as follows. We set the outcome  deterministically to zero if not in the focus population, $Y1^*=0$, or if not screened, $\Delta(a^c)=0$. Otherwise, we  generated the counterfactual outcome as
$Y2(a^c)=\mathbf{1}[U_{Y2}< logit^{-1}(.2+.1A^c+W1+.5W2+2W3+.2E1^c+.2E2^c)]$ with $U_{Y2}\sim Unif(0,1)$. 

Through this process,  we generated each cluster's ``full data'', consisting of the baseline covariates $(E1^c, E2^c, \bm{W1},\bm{W2},\bm{W3})$, the indicators of being in the focus population $\bm{Y1}^*$, 
and the counterfactual indicators of screening and having the outcome $(\bm{\Delta}(a^c),\bm{Y2}(a^c))$ for $a^c \in \{0,1\}$. We converted the full data to the observed data as follows. 
We randomized the trial arm $A^c$ such that $J/2$ clusters were intervention ($A^c=1$) and $J/2$ were control $(A^c=0)$. 
We then assigned the observed indicator for measurement as $\bm{\Delta}= \bm{\Delta}(a^c)$ when $A^c=a^c$, 
the observed indicator for being in the focus population $\bm{Y1} = \bm{Y1}^*\times\bm{\Delta}$, 
and the observed indicator for having the outcome as $\bm{Y2}=\bm{Y2}(a^c)$ when $A^c=a^c$.

We repeated this process for 1000 simulated trials each of $J=\{20,30,50,70\}$ clusters. We calculated the true value of the total effect for a population of 5000 clusters as $\Psi^*=\mathbb{E}[Y^{c*}(1)-Y^{c*}(0)]$ with the cluster-level counterfactual outcome $Y^{c*}(a^c) = \mathbb{P}[Y2(a^c)=1 \mid Y1^*=1]$. We also simulated under the null by setting the coefficients for $a^c$ to be zero in the above data generating process. 

\subsection{Estimators Compared}

For Single-Stage estimators, we implemented GEE among those who screened ($\Delta=1$), adjusting for individual-level and cluster-level covariates ($W1,W2,W3,E1^c, E2^c$), and accounting for clustering with an independent working correlation matrix. To do so, we used the \texttt{MRStdCRT} \texttt{R} package to target the marginal effect on the absolute scale 
\cite{li_model-robust_2025}.

We also implemented Single-Stage TMLE among participants known to be in the focus population ($Y1=1$), adjusting for the same covariates, and accounting for clustering by generating a cluster-level influence curve \cite{schnitzer_effect_2014}.
To do so, we used the \texttt{ltmle} \texttt{R} package with \texttt{id} to indicate cluster membership \cite{schwab_ltmle_2023}.  
While these Single-Stage approaches are more traditional, they condition on the mediator $(\Delta=1)$ or a descendant of the mediator ($Y1=1$) and expected to be biased for the total effect in the underlying focus population. 

Within the Two-Stage framework, we evaluated various approaches for  identification and estimation of the cluster-level outcome $Y^{c*}=\mathbb{P}(Y2=1\mid Y1^*=1)$ in Stage 1. First, we considered   the empirical mean outcome among those measured: $\hat{\mathbb{P}}(Y2=1 \mid \Delta=1)$. We refer to this approach as ``Screened''.
Second, we considered the empirical mean outcome among persons known to be in the focus population and, thus, eligible for the outcome: $\hat{\mathbb{P}}(Y2=1 \mid Y1=1)$. We  refer to this approach as ``Eligible''. While these estimators are intuitive, they do not estimate the cluster-level endpoint $Y^{c*}$ and are expected to be biased for the total effect, even under MCAR (Appendix A of the Supporting Information). 
For Stage 1, we also considered two approaches based on re-expressing the cluster-level outcome $Y^{c*}$  as the ratio: $\mathbb{P}(Y2=1,Y1^*=1) \div \mathbb{P}(Y1^*=1)$. 
In each, the numerator was estimated by taking the empirical proportion with the outcome: $\hat{\mathbb{P}}(Y2=1)$. For the denominator, we considered the MCAR assumption and the unadjusted estimator, calculated as the empirical proportion who are known to be in the focus population: $\hat{\mathbb{P}}(Y1=1 \mid \Delta=1)$. 
For the denominator, we also considered the MAR assumption for identification and used TMLE to  estimate $\mathbb{E}[\mathbb{E}(Y1 \mid \Delta=1, W)]$,  adjusting for differences in covariate values $W$ between persons screened versus not screened. 
For these two ratio-based approaches, we divided  the estimated numerator by the estimated denominator to obtain the cluster-level endpoint estimate $\hat{Y}^c$.
When evaluating the total effect in Stage 2, we implemented the unadjusted effect estimator for all approaches. Specifically, we took the difference in the empirical mean of the cluster-level endpoints $\hat{Y}^c$ by arm.
For the approach using TMLE in Stage 1, we implemented TMLE with Adaptive Pre-specification in Stage 2. Specifically, we used cross-validation to select among working GLMs adjusting for at most 1 cluster-level covariate.

\subsection{Results}

In these simulations, the true value of the total effect was 4.28\%.
For $J=\{20,30,50,70\}$ clusters with an average size of 200 persons, Table 1 provides the following metrics across 1000 iterations: the average of the point estimate and  95\% confidence intervals, bias (average deviation of point estimates from the true effect), the Monte Carlo standard deviation, the average standard error estimate, coverage (proportion of 95\% confidence intervals containing the true effect), and attained power  (proportion of tests rejecting the null hypothesis).

As expected, both Single-Stage approaches were meaningfully biased towards the null. Specifically, both GEE and the Single-Stage TMLE underestimated the total effect by conditioning on a mediator (screening) or a descendant of the mediator (eligibility). For both estimators, the resulting confidence interval coverage was much less than the nominal rate of 95\% and decreased with increasing sample size. Indeed, at a sample size of $J=70$,  coverage was 52.5\% for GEE and was $53.5\%$ for the Single-Stage TMLE. 

Within the Two-Stage framework, the two approaches estimating the cluster-level endpoint $\hat{Y}^c$ among those screened $\hat{\mathbb{P}}(Y2=1\mid \Delta=1)$ and among those eligible $\hat{\mathbb{P}}(Y2=1\mid Y1=1)$ were also meaningfully biased. This was expected, because both did not target the correct estimand in Stage 1 (Appendix A of the Supporting Information). The ``Screened'' approach was biased toward the null. Its 95\% confidence interval coverage  ranged from 81.3\% for the smallest sample size to 52.4\% for the largest sample size.  On average, the ``Eligible'' approach generated a negative effect estimate --- suggesting the positive intervention was actually harmful. Its 95\% confidence interval coverage ranged from 62.5\% for the smallest sample size to 13.5\% for the largest sample size.

\begin{table}[!ht]
\centering
\begin{threeparttable}
\caption{Performance of Single-Stage and Two-Stage approaches across 1000 simulated trials with $J=\{20,30,50,70\}$ clusters and a true effect of 4.28\%.}
\label{tab:simulation_results}
\begin{tabular}{p{4.75cm} c l c c c c c}
\hline
  \textbf{Method} & \textbf{J} & \textbf{Pt (95\% CI)} & \textbf{Bias} & \boldmath$\hat{\sigma}$ & \boldmath$\sigma$ & \textbf{Coverage} & \textbf{Power} \\ 
 \hline
\emph{Single-Stage Methods} \\
\multirow[t]{4}{=}{GEE among screened ($\Delta=1$) adjusting for ($W,E^c$)}
 &  20 & $0.89\;(-9.61,\;11.38)$ & -3.40 & 0.0411 & 0.0502 & 91.5 & 3.3 \\ 
   &  30 & $0.74\;(-6.10,\;7.59)$ & -3.54 & 0.0300 & 0.0335 & 83.6 & 6.2 \\ 
   &  50 & $0.89\;(-3.73,\;5.50)$ & -3.40 & 0.0212 & 0.0230 & 69.7 & 5.5 \\ 
   &  70 & $0.73\;(-2.92,\;4.37)$ & -3.56 & 0.0174 & 0.0183 & 52.2 & 5.9 \\ 
\multirow[t]{4}{=}{TMLE among eligible ($Y1=1$) adjusting for ($W,E^c$)}
   &  20 & $1.58\;(-3.51,\;6.67)$ & -2.71 & 0.0285 & 0.0243 & 78.9 & 14.1 \\ 
   &  30 & $1.56\;(-2.74,\;5.85)$ & -2.73 & 0.0234 & 0.0210 & 72.2 & 12.8 \\ 
   &  50 & $1.67\;(-1.63,\;4.98)$ & -2.61 & 0.0174 & 0.0164 & 66.6 & 16.9 \\ 
   &  70 & $1.57\;(-1.21,\;4.35)$ & -2.71 & 0.0141 & 0.0139 & 53.5 & 19.6 \\ 
 \emph{Two-Stage Methods} \\
 \multirow[t]{4}{=}{``Screened'' in Stage1 \& Unadjusted for Stage2}
 &  20 & $0.48\;(-6.82,\;7.77)$ & -3.81 & 0.0356 & 0.0347 & 81.3 & 5.6 \\ 
   &  30 & $0.69\;(-5.22,\;6.59)$ & -3.60 & 0.0303 & 0.0288 & 76.6 & 6.4 \\ 
   &  50 & $0.78\;(-3.62,\;5.17)$ & -3.50 & 0.0224 & 0.0224 & 64.0 & 6.5 \\ 
   &  70 & $0.70\;(-3.03,\;4.44)$ & -3.58 & 0.0194 & 0.0191 & 52.4 & 7.9 \\ 
\multirow[t]{4}{=}{``Eligible'' in Stage1 \& Unadjusted for Stage2}
 &  20 & $-1.86\;(-9.30,\;5.58)$ & -6.14 & 0.0363 & 0.0354 & 62.5 & 8.5 \\ 
   &  30 & $-1.77\;(-7.78,\;4.25)$ & -6.05 & 0.0309 & 0.0294 & 48.8 & 9.6 \\ 
   &  50 & $-1.65\;(-6.17,\;2.88)$ & -5.93 & 0.0239 & 0.0231 & 28.7 & 12.7 \\ 
   &  70 & $-1.73\;(-5.55,\;2.10)$ & -6.01 & 0.0192 & 0.0195 & 13.5 & 14.0 \\  
\multirow[t]{4}{=}{Unadjusted ratio in Stage1 \& Unadjusted for Stage2}
 &  20 & $2.59\;(-2.00,\;7.17)$ & -1.70 & 0.0222 & 0.0218 & 89.3 & 20.4 \\ 
  &  30 & $2.68\;(-1.00,\;6.36)$ & -1.60 & 0.0195 & 0.0180 & 83.7 & 30.9 \\ 
  &  50 & $2.81\;(0.03,\;5.59)$ & -1.47 & 0.0144 & 0.0142 & 81.3 & 50.0 \\ 
  &  70 & $2.73\;(0.38,\;5.08)$ & -1.56 & 0.0120 & 0.0120 & 74.8 & 62.2 \\ 
 
\multirow[t]{4}{=}{TMLE ratio in Stage1 \& Unadjusted for Stage2}
 &  20 & $4.10\;(-0.74,\;8.93)$ & -0.19 & 0.0236 & 0.0230 & 94.5 & 37.7 \\ 
 &  30 & $4.21\;(0.32,\;8.09)$ & -0.08 & 0.0205 & 0.0190 & 93.6 & 55.4 \\ 
 &  50 & $4.32\;(1.39,\;7.26)$ & 0.04 & 0.0151 & 0.0150 & 94.7 & 81.8 \\ 
 &  70 & $4.26\;(1.78,\;6.73)$ & -0.03 & 0.0126 & 0.0126 & 94.0 & 92.3 \\
\multirow[t]{4}{=}{TMLE ratio  in Stage1 \& TMLE for Stage2}
 &  20 & $4.16\;(-0.63,\;8.94)$ & -0.13 & 0.0216 & 0.0228 & 96.2 & 40.5 \\ 
   &  30 & $4.17\;(0.54,\;7.81)$ & -0.11 & 0.0171 & 0.0177 & 96.0 & 60.8 \\ 
   &  50 & $4.30\;(1.65,\;6.95)$ & 0.01 & 0.0126 & 0.0135 & 95.6 & 90.1 \\ 
   &  70 & $4.23\;(2.05,\;6.41)$ & -0.05 & 0.0108 & 0.0111 & 95.4 & 96.7 \\ 
     \hline
\end{tabular}
\begin{tablenotes}
  \footnotesize
  \item Pt (95\% CI): average of the point estimate and  95\% confidence intervals (in \%)
  \item Bias: average deviation of point estimates from the true effect (in \%)
  \item $\hat{\sigma}$: average estimate of the standard error
  \item $\sigma$: standard deviation of the point estimates
  \item Coverage: Proportion of 95\% confidence interval containing the true effect (in \%)
  \item Power: Proportion of trials correctly rejecting the false null hypothesis (in \%)
  \item ``Screened'': $\hat{Y}^c=\hat{\mathbb{P}}(Y2=1 \mid \Delta=1)$ 
  \item ``Eligible'':  $\hat{Y}^c=\hat{\mathbb{P}}(Y2=1 \mid Y1=1)$
  \item Unadjusted ratio: $\hat{Y}^c=\hat{\mathbb{P}}(Y2=1)$ $\div$ $\hat{\mathbb{P}}(Y1=1 \mid \Delta=1)$
  \item TMLE ratio: $\hat{Y}^c=\hat{\mathbb{P}}(Y2=1)$ $\div$ $\hat{\mathbb{E}}[\hat{\mathbb{E}}(Y1 \mid \Delta=1, W)]$
\end{tablenotes}
\end{threeparttable}
\end{table}

As expected, the performance of the Two-Stage approaches was improved when taking the ratio-based approach to Stage 1. Also as expected, the unadjusted approach to estimating the cluster-level endpoint $\hat{Y}^c$ as $\hat{\mathbb{P}}(Y2=1)\div \hat{\mathbb{P}}(Y1=1 \mid \Delta=1)$ was still biased, because it failed to account for differences between those screened versus not screened. Its confidence interval coverage remained well-below the target and was in the range of 74.8\% to 89.3\%.  
In contrast, relaxing the missing data assumption and using TMLE in Stage 1 resulted in negligible bias and nominal confidence interval coverage across all sample sizes. Additionally, using TMLE in Stage 2 was more  efficient, as expected. Indeed, adaptive selection of the cluster-level adjustment covariates resulted in 2.8\% to 8.3\% higher power, while maintaining 95\% confidence interval coverage. Furthermore, as shown in Appendix B of the Supporting Information, the approach maintained strict Type-I error control under the null.

\section{Extensions}
\label{sec:extensions}

We now present an extension of the above methods to (1) accommodate time-varying covariates and (2) allow the exposure to impact the focus population. Figure 3 provides an updated causal graph, where  $\bm{L}$ denotes the matrix of post-baseline, individual-level covariates. In OPAL, these covariates could include measures of mobility and alcohol use. As shown in Figure 3, there are now several pathways of the intervention effect. As before, the intervention $A^c$ directly impacts outcomes $\bm{Y2}$ and indirectly  through its effect on health screening $\bm{\Delta}$. Additionally, the intervention now indirectly impacts outcomes through its effect on the time-varying covariates $\bm{L}$ and on underlying status defining who is in the focus population $\bm{Y1}^*$. Importantly, the time-varying covariates $\bm{L}$ effectively confound the relationship between health screening $\bm{\Delta}$ and health outcomes $\bm{Y2}$. The corresponding non-parametric structural equations are provided in  Appendix C in the Supporting Information.

\begin{figure}[H]
   \centering
   \includegraphics[width=0.7\textwidth]{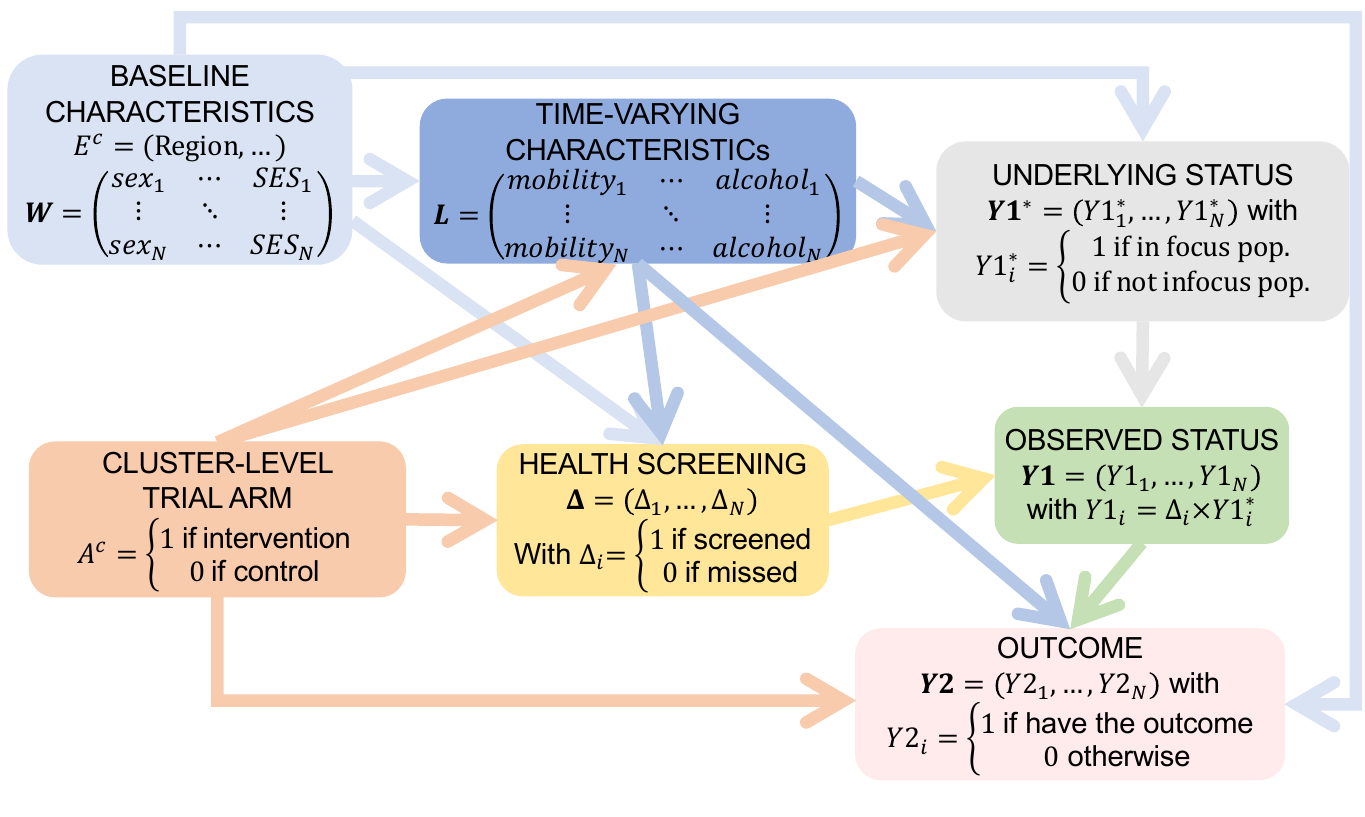}
    \caption{A simplified hierarchical causal graph to illustrate the data generating process where the cluster-level intervention $A^c$ impacts time-varying covariates $\bm{L}$ as well as underlying status $\bm{Y1}^*$. The time-varying covariates $\bm{L}$ also impact measurement $\bm{\Delta}$, underlying status  $\bm{Y1}^*$, and the outcome  $\bm{Y2}$.
    For simplicity, we have again omitted the unmeasured factors: $\bm{U}=(U_{E^c}, U_{\bm{W}}, U_{A^c},U_{\bm{L}}, U_{\bm{Y1}^*}, U_{\bm{\Delta}},  U_{\bm{Y2}})$.}
   \label{fig:ext_dag}
\end{figure}

Our interest remains in the total effect $
\Psi^*=\mathbb{E}[Y^{c*} (1)-Y^{c*} (0)]$.
However, the focus population is now a function of the intervention strategy:
$ Y^{c*}(a^c)=\mathbb{P}[Y2(a^c)=1 \mid Y1^*(a^c)=1]$. Despite a more complicated causal model and causal estimand, the Two-Stage approach is still applicable.
In Stage 1, stratification on each cluster obviates the multilevel-mediation-missing data problem as well as the missing-data equivalent to time-dependent confounding \cite{balzer_two-stage_2023}.  
Within each cluster the cluster-level variables $(E^c, A^c)$ are fixed. Therefore, our Stage 1 causal estimand remains the same and is the counterfactual probability of the outcome for persons in the underlying focus population:
\[
Y^{c*}= \mathbb{P}(Y2=1\mid Y1^*=1) = \frac{\mathbb{P}(Y2=1, Y1^*=1)}{\mathbb{P}(Y1^*=1)}
\]
As before, the numerator is trivially identified as the  proportion with the outcome: $\mathbb{P}(Y2=1)$. 

Now, however, we can account for baseline and time-varying covariates when identifying and estimating the denominator.
Specifically, we assume that within values of baseline and post-baseline covariates, health status among those screened is representative among health status among those missed: 
$ Y1^* \perp \Delta \mid  W,L $. Additionally, we require that the probability of being measured within covariate values is bounded away from 0:
$ P( \Delta =1 \mid W=w,L=l) >0 \ \text{ for all possible } w,l $. Under these assumptions, we can write the statistical estimand for the denominator as 
$\mathbb{E}[\mathbb{E}(Y1 \mid \Delta=1, W,L)]$. Estimation for the denominator proceeds as before using TMLE with a larger adjustment set. Then we combine the numerator and denominator estimates to generate $\hat{Y}^c$, repeat for all $J$ clusters, and conduct Stage 2 estimation and inference using TMLE with Adaptive Pre-specification. 
 We emphasize that Stage 1's identification and estimation of the cluster-level endpoint $Y^c = \mathbb{P}(Y2=1) \div \mathbb{E}[\mathbb{E}(Y1 \mid \Delta=1, W,L)]$ is done in each cluster separately. This allows the missingness mechanism to vary by cluster and facilitates our adjustment for time-varying covariates impacted by the trial arm.

\section{Discussion}

Novel challenges and opportunities arise when health strategies aim to expand reach into the focus population and improve outcomes. 
Specifically, the strategy improves health outcomes directly and indirectly through expanded reach, while outcomes are only measured among those reached.
In this common scenario, existing methods to address differential outcome measurement would effectively block the indirect effect and bias the total effect towards the null. Building on prior work \cite{balzer_two-stage_2023}, we used causal models to specify the data generating process and define the causal estimand with Counterfactual Strata Effects  capturing the intervention effect in the underlying target population, without enforcing 100\% compliance on screening. Then we extended the Two-Stage TMLE framework for estimation and inference.
Our simulations demonstrated that our proposed approach obtained nominal confidence interval coverage, while more traditional approaches failed.  We plan to use our novel approach in the primary, pre-specified analysis of the OPAL trial, which is evaluating the effectiveness of community-based, recruitment strategies on the use of biomedical HIV prevention among adults at alcohol-serving venues in Kenya and Uganda.

This work has limitations, and future work is needed. First, the Two-Stage TMLE framework obviates the multilevel-mediation-missing data problem by fully stratifying on each cluster in Stage 1. This can lead to poor data support when there are small clusters and/or a large adjustment set. We are exploring strategies to adaptively pool across clusters within arm in Stage 1  \cite{balzer_SER_2025}. 
We note, however, that pooling approaches would require homogeneity assumptions that are not needed in our Two-Stage framework.
Second, to match the motivating example, we focused on a binary outcome. However, our methods are equally applicable to other types of outcomes. Also to match the motivating example, we studied cluster randomized trials, which randomize groups to the intervention and comparator strategy. We plan to extend this work to observational studies where the group-level strategy is delivered in a non-randomized strategy. In such a setting, we would need to adjust for cluster-level confounding in Stage 2 of Two-Stage TMLE. We also plan to extend this work to individually randomized trials where the health promotion strategy is delivered at the individual-level. Two-Stage TMLE, designed for hierarchical data, would not be immediately applicable in such a setting; however, we hypothesize that creating artificial clusters within arm and implementing the Two-Stage procedure (with appropriate weights) should facilitate estimation of the total effect, despite the mediation-missing data problem. Finally, in this paper as well as previous work with Counterfactual Strata Effects \cite{balzer_evaluation_2017,balzer_far_2020, balzer_two-stage_2023, nugent_blurring_2023, gupta_mechanism_2024,petersen_causal_2024}, we have estimated the numerator and denominator separately prior to taking their ratio. The denominator defines being in the focus population of interest and is always bigger than the numerator. Therefore, the ratio is guaranteed to be bounded between 0 and 1.  Using an efficient estimator for both the numerator and denominator implies via the Delta Method that the ratio estimator  is asymptotically linear with influence function equal to the canonical gradient of the ratio parameter and, thus, achieves the semi-parametric efficiency bound \cite{vanderVaart1998,Tsiatis2006}. Nonetheless, it is of interest to develop a TMLE directly targeting the ratio. 
Future work is warranted.

\section{Acknowledgments}
We gratefully acknowledge the participants and communities for taking part in the OPAL study. We also gratefully acknowledge the larger OPAL study team and the SEARCH collaboration. We  thank Dr. Diane Havlir and Dr. Maya L. Petersen, who together with Dr. Moses R. Kamya are
the MPIs of the SEARCH collaboration. We finally thank Dr. Mark van der Laan for his advice on this work.

\section{Conflict of interest}
The authors declare no potential conflict of interests.

\section{Funding source}
This work was supported, in part, by the National Institutes of Health under awards: R01AA030464, K24AA031211, U01AI150510 and R01MH140685. The content is solely the responsibility of the authors and does not necessarily represent the official views of the National Institutes of Health.

\pagestyle{headings}

\newpage

\bibliographystyle{unsrtnat}
\bibliography{main}

@article{balzer2025machinelearningoptimizeprecision,
      title={Machine learning to optimize precision in the analysis of randomized trials: A journey in pre-specified, yet data-adaptive learning}, 
      author={Laura B. Balzer and Mark J. van der Laan and Maya L. Petersen},
      year={2026},
    number = {In Press},
	journal = {Clinical Trials},
      url={https://arxiv.org/abs/2512.13610}, 
}

@article{van_der_laan_super_2007,
	title = {Super learner},
	volume = {6},
	issn = {1544-6115},
	doi = {10.2202/1544-6115.1309},
	language = {eng},
	journal = {Statistical Applications in Genetics and Molecular Biology},
	author = {van der Laan, Mark J. and Polley, Eric C. and Hubbard, Alan E.},
	year = {2007},
	pmid = {17910531},
	pages = {Article25},
}

@article{balzer_two-stage_2023,
	title = {Two-{Stage} {TMLE} to reduce bias and improve efficiency in cluster randomized trials},
	volume = {24},
	issn = {1465-4644, 1468-4357},
	url = {https://academic.oup.com/biostatistics/article/24/2/502/6481160},
	doi = {10.1093/biostatistics/kxab043},
	language = {en},
	number = {2},
	urldate = {2024-02-28},
	journal = {Biostatistics},
	author = {Balzer, Laura B and Van Der Laan, Mark and Ayieko, James and Kamya, Moses and Chamie, Gabriel and Schwab, Joshua and Havlir, Diane V and Petersen, Maya L},
	month = apr,
	year = {2023},
	pages = {502--517},
}

@article{moore_covariate_2009,
	title = {Covariate adjustment in randomized trials with binary outcomes: {Targeted} maximum likelihood estimation},
	volume = {28},
	issn = {0277-6715},
	shorttitle = {Covariate adjustment in randomized trials with binary outcomes},
	url = {https://www.ncbi.nlm.nih.gov/pmc/articles/PMC2857590/},
	doi = {10.1002/sim.3445},
	number = {1},
	urldate = {2024-03-28},
	journal = {Statistics in medicine},
	author = {Moore, K. L. and van der Laan, M. J.},
	month = jan,
	year = {2009},
	pmid = {18985634},
	pmcid = {PMC2857590},
	pages = {39--64},
}

@inproceedings{didelez_direct_2006,
	address = {Arlington, Virginia, USA},
	series = {{UAI}'06},
	title = {Direct and indirect effects of sequential treatments},
	isbn = {978-0-9749039-2-7},
	urldate = {2024-05-07},
	booktitle = {Proceedings of the {Twenty}-{Second} {Conference} on {Uncertainty} in {Artificial} {Intelligence}},
	publisher = {AUAI Press},
	author = {Didelez, Vanessa and Dawid, A. Philip and Geneletti, Sara},
	month = jul,
	year = {2006},
	pages = {138--146},
}

@article{vanderweele_mediation_2017,
	title = {Mediation analysis with time varying exposures and mediators},
	volume = {79},
	issn = {1369-7412},
	doi = {10.1111/rssb.12194},
	language = {eng},
	number = {3},
	journal = {Journal of the Royal Statistical Society. Series B, Statistical Methodology},
	author = {VanderWeele, Tyler J. and Tchetgen Tchetgen, Eric J.},
	month = jun,
	year = {2017},
	pmid = {28824285},
	pmcid = {PMC5560424},
	pages = {917--938},
}

@article{rudolph_robust_2018,
	title = {Robust and {Flexible} {Estimation} of {Stochastic} {Mediation} {Effects}: {A} {Proposed} {Method} and {Example} in a {Randomized} {Trial} {Setting}},
	volume = {7},
	issn = {2194-9263},
	shorttitle = {Robust and {Flexible} {Estimation} of {Stochastic} {Mediation} {Effects}},
	url = {https://www.ncbi.nlm.nih.gov/pmc/articles/PMC8136358/},
	doi = {10.1515/em-2017-0007},
	number = {1},
	urldate = {2024-05-07},
	journal = {Epidemiologic methods},
	author = {Rudolph, Kara E. and Sofrygin, Oleg and Zheng, Wenjing and van der Laan, Mark J.},
	year = {2018},
	pmid = {34026421},
	pmcid = {PMC8136358},
	pages = {20170007},
}

@article{mackinnon_mediation_2007,
	title = {Mediation {Analysis}},
	volume = {58},
	issn = {0066-4308},
	url = {https://www.ncbi.nlm.nih.gov/pmc/articles/PMC2819368/},
	doi = {10.1146/annurev.psych.58.110405.085542},
	urldate = {2024-05-07},
	journal = {Annual review of psychology},
	author = {MacKinnon, David P. and Fairchild, Amanda J. and Fritz, Matthew S.},
	year = {2007},
	pmid = {16968208},
	pmcid = {PMC2819368},
	pages = {593},
}

@article{rubin_inference_1976,
	title = {Inference and missing data},
	volume = {63},
	issn = {0006-3444},
	url = {https://doi.org/10.1093/biomet/63.3.581},
	doi = {10.1093/biomet/63.3.581},
	number = {3},
	urldate = {2024-05-07},
	journal = {Biometrika},
	author = {RUBIN, DONALD B.},
	month = dec,
	year = {1976},
	pages = {581--592},
}

@book{vaart_asymptotic_1998,
	edition = {1},
	title = {Asymptotic {Statistics}},
	copyright = {https://www.cambridge.org/core/terms},
	isbn = {978-0-511-80225-6 978-0-521-49603-2 978-0-521-78450-4},
	url = {https://www.cambridge.org/core/product/identifier/9780511802256/type/book},
	language = {en},
	urldate = {2024-05-07},
	publisher = {Cambridge University Press},
	author = {Vaart, A. W. Van Der},
	month = oct,
	year = {1998},
	doi = {10.1017/CBO9780511802256},
}

@article{balzer_far_2020,
	title = {Far from {MCAR}: {Obtaining} population-level estimates of {HIV} viral suppression},
	volume = {31},
	issn = {1044-3983},
	shorttitle = {Far from {MCAR}},
	url = {https://www.ncbi.nlm.nih.gov/pmc/articles/PMC8105880/},
	doi = {10.1097/EDE.0000000000001215},
	number = {5},
	urldate = {2024-05-09},
	journal = {Epidemiology (Cambridge, Mass.)},
	author = {Balzer, Laura B. and Ayieko, James and Kwarisiima, Dalsone and Chamie, Gabriel and Charlebois, Edwin D. and Schwab, Joshua and van der Laan, Mark J. and Kamya, Moses R. and Havlir, Diane V. and Petersen, Maya L.},
	month = sep,
	year = {2020},
	pmid = {32452912},
	pmcid = {PMC8105880},
	pages = {620--627},
}

@book{van_der_laan_targeted_2011,
	address = {New York, NY},
	series = {Springer {Series} in {Statistics}},
	title = {Targeted {Learning}: {Causal} {Inference} for {Observational} and {Experimental} {Data}},
	copyright = {https://www.springernature.com/gp/researchers/text-and-data-mining},
	isbn = {978-1-4419-9781-4 978-1-4419-9782-1},
	shorttitle = {Targeted {Learning}},
	url = {https://link.springer.com/10.1007/978-1-4419-9782-1},
	language = {en},
	urldate = {2024-05-10},
	publisher = {Springer},
	author = {Van Der Laan, Mark J. and Rose, Sherri},
	year = {2011},
	doi = {10.1007/978-1-4419-9782-1},
}

@article{balzer_adaptivePre_2016,
	title = {Adaptive {Pre}-specification in {Randomized} {Trials} {With} and {Without} {Pair}-{Matching}},
	volume = {35},
	issn = {0277-6715},
	url = {https://www.ncbi.nlm.nih.gov/pmc/articles/PMC5084457/},
	doi = {10.1002/sim.7023},
	number = {25},
	urldate = {2024-06-01},
	journal = {Statistics in medicine},
	author = {Balzer, Laura B. and van der Laan, Mark J. and Petersen, Maya L.},
	month = nov,
	year = {2016},
	pmid = {27436797},
	pmcid = {PMC5084457},
	pages = {4528--4545},
}

@article{young_causal_2020,
	title = {A causal framework for classical statistical estimands in failure-time settings with competing events},
	volume = {39},
	issn = {1097-0258},
	doi = {10.1002/sim.8471},
	language = {eng},
	number = {8},
	journal = {Statistics in Medicine},
	author = {Young, Jessica G. and Stensrud, Mats J. and Tchetgen Tchetgen, Eric J. and Hernán, Miguel A.},
	month = apr,
	year = {2020},
	pmid = {31985089},
	pmcid = {PMC7811594},
	pages = {1199--1236},
}

@article{greenland_causal_1999,
	title = {Causal diagrams for epidemiologic research},
	volume = {10},
	issn = {1044-3983},
	language = {eng},
	number = {1},
	journal = {Epidemiology (Cambridge, Mass.)},
	author = {Greenland, S. and Pearl, J. and Robins, J. M.},
	month = jan,
	year = {1999},
	pmid = {9888278},
	pages = {37--48},
}

@article{balzer_new_2019,
	title = {A new approach to hierarchical data analysis: {Targeted} maximum likelihood estimation for the causal effect of a cluster-level exposure},
	volume = {28},
	issn = {0962-2802},
	shorttitle = {A new approach to hierarchical data analysis},
	url = {https://www.ncbi.nlm.nih.gov/pmc/articles/PMC6173669/},
	doi = {10.1177/0962280218774936},
	number = {6},
	urldate = {2024-07-02},
	journal = {Statistical methods in medical research},
	author = {Balzer, Laura B and Zheng, Wenjing and van der Laan, Mark J and Petersen, Maya L},
	month = jun,
	year = {2019},
	pmid = {29921160},
	pmcid = {PMC6173669},
	pages = {1761--1780},
}

@book{pearl_causality_2009,
	address = {Cambridge},
	edition = {2},
	title = {Causality},
	isbn = {978-0-521-89560-6},
	url = {https://www.cambridge.org/core/books/causality/B0046844FAE10CBF274D4ACBDAEB5F5B},
	urldate = {2024-07-03},
	publisher = {Cambridge University Press},
	author = {Pearl, Judea},
	year = {2009},
	doi = {10.1017/CBO9780511803161},
}

@article{balzer_adapt_2024,
	title = {Adaptive selection of the optimal strategy to improve precision and power in randomized trials},
	volume = {80},
	issn = {0006-341X},
	url = {https://doi.org/10.1093/biomtc/ujad034},
	doi = {10.1093/biomtc/ujad034},
	number = {1},
	urldate = {2024-07-03},
	journal = {Biometrics},
	author = {Balzer, Laura B and Cai, Erica and Godoy Garraza, Lucas and Amaranath, Pracheta},
	month = mar,
	year = {2024},
	pages = {ujad034},
}

@article{stitelman_collaborative_2010,
	title = {Collaborative targeted maximum likelihood for time to event data},
	volume = {6},
	issn = {1557-4679},
	doi = {10.2202/1557-4679.1249},
	language = {eng},
	number = {1},
	journal = {The International Journal of Biostatistics},
	author = {Stitelman, Ori M. and van der Laan, Mark J.},
	year = {2010},
	pmid = {21969976},
	pages = {Article 21},
}

@article{frangakis_principal_2002,
	title = {Principal {Stratification} in {Causal} {Inference}},
	volume = {58},
	issn = {0006-341X},
	url = {https://www.ncbi.nlm.nih.gov/pmc/articles/PMC4137767/},
	number = {1},
	urldate = {2024-07-05},
	journal = {Biometrics},
	author = {Frangakis, Constantine E. and Rubin, Donald B.},
	month = mar,
	year = {2002},
	pmid = {11890317},
	pmcid = {PMC4137767},
	pages = {21--29},
}

@article{grilli_nonparametric_2008-1,
	title = {Nonparametric {Bounds} on the {Causal} {Effect} of {University} {Studies} on {Job} {Opportunities} {Using} {Principal} {Stratification}},
	volume = {33},
	issn = {1076-9986},
	url = {https://doi.org/10.3102/1076998607302627},
	doi = {10.3102/1076998607302627},
	language = {en},
	number = {1},
	urldate = {2024-07-05},
	journal = {Journal of Educational and Behavioral Statistics},
	author = {Grilli, Leonardo and Mealli, Fabrizia},
	month = mar,
	year = {2008},
	note = {Publisher: American Educational Research Association},
	pages = {111--130},
}

@article{janvin_causal_2024-1,
	title = {Causal inference with recurrent and competing events},
	volume = {30},
	issn = {1572-9249},
	url = {https://doi.org/10.1007/s10985-023-09594-8},
	doi = {10.1007/s10985-023-09594-8},
	language = {en},
	number = {1},
	urldate = {2024-07-07},
	journal = {Lifetime Data Analysis},
	author = {Janvin, Matias and Young, Jessica G. and Ryalen, Pål C. and Stensrud, Mats J.},
	month = jan,
	year = {2024},
	pages = {59--118},
}

@article{petersen_estimation_2006,
	title = {Estimation of direct causal effects},
	volume = {17},
	issn = {1044-3983},
	doi = {10.1097/01.ede.0000208475.99429.2d},
	language = {eng},
	number = {3},
	journal = {Epidemiology (Cambridge, Mass.)},
	author = {Petersen, Maya L. and Sinisi, Sandra E. and van der Laan, Mark J.},
	month = may,
	year = {2006},
	pmid = {16617276},
	pages = {276--284},
}

@article{austin_introduction_2016,
	title = {Introduction to the {Analysis} of {Survival} {Data} in the {Presence} of {Competing} {Risks}},
	volume = {133},
	issn = {1524-4539},
	doi = {10.1161/CIRCULATIONAHA.115.017719},
	language = {eng},
	number = {6},
	journal = {Circulation},
	author = {Austin, Peter C. and Lee, Douglas S. and Fine, Jason P.},
	month = feb,
	year = {2016},
	pmid = {26858290},
	pmcid = {PMC4741409},
	pages = {601--609},
}

@article{rudolph_causal_2019,
	title = {Causal {Mediation} {Analysis} {With} {Observational} {Data}: {Considerations} and {Illustration} {Examining} {Mechanisms} {Linking} {Neighborhood} {Poverty} to {Adolescent} {Substance} {Use}},
	volume = {188},
	issn = {0002-9262},
	shorttitle = {Causal {Mediation} {Analysis} {With} {Observational} {Data}},
	url = {https://doi.org/10.1093/aje/kwy248},
	doi = {10.1093/aje/kwy248},
	number = {3},
	urldate = {2024-07-11},
	journal = {American Journal of Epidemiology},
	author = {Rudolph, Kara E and Goin, Dana E and Paksarian, Diana and Crowder, Rebecca and Merikangas, Kathleen R and Stuart, Elizabeth A},
	month = mar,
	year = {2019},
	pages = {598--608},
}

@misc{petersen_causal_2024,
	address = {Copenhagen, Denmark},
	title = {The {Causal} {Roadmap} in the {Age} of {AI}: {From} {All} {Wheel} {Drive} to {Formula} 1},
	author = {Petersen, Maya},
	month = apr,
	year = {2024},
}

@article{petersen_association_2017,
	title = {Association of {Implementation} of a {Universal} {Testing} and {Treatment} {Intervention} {With} {HIV} {Diagnosis}, {Receipt} of {Antiretroviral} {Therapy}, and {Viral} {Suppression} in {East} {Africa}},
	volume = {317},
	issn = {1538-3598},
	doi = {10.1001/jama.2017.5705},
	number = {21},
	journal = {JAMA},
	author = {Petersen, Maya and Balzer, Laura and Kwarsiima, Dalsone and Sang, Norton and Chamie, Gabriel and Ayieko, James and Kabami, Jane and Owaraganise, Asiphas and Liegler, Teri and Mwangwa, Florence and Kadede, Kevin and Jain, Vivek and Plenty, Albert and Brown, Lillian and Lavoy, Geoff and Schwab, Joshua and Black, Douglas and van der Laan, Mark and Bukusi, Elizabeth A. and Cohen, Craig R. and Clark, Tamara D. and Charlebois, Edwin and Kamya, Moses and Havlir, Diane},
	month = jun,
	year = {2017},
	pmid = {28586888},
	pmcid = {PMC5734234},
	pages = {2196--2206},
}

@article{havlir_hiv_2019,
	title = {{HIV} {Testing} and {Treatment} with the {Use} of a {Community} {Health} {Approach} in {Rural} {Africa}},
	volume = {381},
	issn = {0028-4793},
	url = {https://www.nejm.org/doi/full/10.1056/NEJMoa1809866},
	doi = {10.1056/NEJMoa1809866},
	number = {3},
	urldate = {2024-07-11},
	journal = {New England Journal of Medicine},
	author = {Havlir, Diane V. and Balzer, Laura B. and Charlebois, Edwin D. and Clark, Tamara D. and Kwarisiima, Dalsone and Ayieko, James and Kabami, Jane and Sang, Norton and Liegler, Teri and Chamie, Gabriel and Camlin, Carol S. and Jain, Vivek and Kadede, Kevin and Atukunda, Mucunguzi and Ruel, Theodore and Shade, Starley B. and Ssemmondo, Emmanuel and Byonanebye, Dathan M. and Mwangwa, Florence and Owaraganise, Asiphas and Olilo, Winter and Black, Douglas and Snyman, Katherine and Burger, Rachel and Getahun, Monica and Achando, Jackson and Awuonda, Benard and Nakato, Hellen and Kironde, Joel and Okiror, Samuel and Thirumurthy, Harsha and Koss, Catherine and Brown, Lillian and Marquez, Carina and Schwab, Joshua and Lavoy, Geoff and Plenty, Albert and Wafula, Erick Mugoma and Omanya, Patrick and Chen, Yea-Hung and Rooney, James F. and Bacon, Melanie and Laan, Mark van der and Cohen, Craig R. and Bukusi, Elizabeth and Kamya, Moses R. and Petersen, Maya},
	month = jul,
	year = {2019},
	note = {Publisher: Massachusetts Medical Society
\_eprint: https://www.nejm.org/doi/pdf/10.1056/NEJMoa1809866},
	pages = {219--229},
}

@article{nugent_blurring_2023,
	title = {Blurring cluster randomized trials and observational studies: {Two}-{Stage} {TMLE} for subsampling, missingness, and few independent units},
	issn = {1465-4644},
	shorttitle = {Blurring cluster randomized trials and observational studies},
	url = {https://doi.org/10.1093/biostatistics/kxad015},
	doi = {10.1093/biostatistics/kxad015},
	urldate = {2024-07-11},
	journal = {Biostatistics},
	author = {Nugent, Joshua R and Marquez, Carina and Charlebois, Edwin D and Abbott, Rachel and Balzer, Laura B},
	month = aug,
	year = {2023},
	pages = {kxad015},
}

@article{abbott_incident_2024,
	title = {Incident {Tuberculosis} {Infection} {Is} {Associated} {With} {Alcohol} {Use} in {Adults} in {Rural} {Uganda}},
	issn = {1058-4838},
	url = {https://doi.org/10.1093/cid/ciae304},
	doi = {10.1093/cid/ciae304},
	urldate = {2024-07-11},
	journal = {Clinical Infectious Diseases},
	author = {Abbott, Rachel and Landsiedel, Kirsten and Atukunda, Mucunguzi and Puryear, Sarah B and Chamie, Gabriel and Hahn, Judith A and Mwangwa, Florence and Kakande, Elijah and Petersen, Maya L and Havlir, Diane V and Charlebois, Edwin and Balzer, Laura B and Kamya, Moses R and Marquez, Carina},
	month = jun,
	year = {2024},
	pages = {ciae304},
}

@article{marquez_community-wide_2024,
	title = {Community-{Wide} {Universal} {HIV} {Test} and {Treat} {Intervention} {Reduces} {Tuberculosis} {Transmission} in {Rural} {Uganda}: {A} {Cluster}-{Randomized} {Trial}},
	volume = {78},
	issn = {1058-4838},
	shorttitle = {Community-{Wide} {Universal} {HIV} {Test} and {Treat} {Intervention} {Reduces} {Tuberculosis} {Transmission} in {Rural} {Uganda}},
	url = {https://doi.org/10.1093/cid/ciad776},
	doi = {10.1093/cid/ciad776},
	number = {6},
	urldate = {2024-07-11},
	journal = {Clinical Infectious Diseases},
	author = {Marquez, Carina and Atukunda, Mucunguzi and Nugent, Joshua and Charlebois, Edwin D and Chamie, Gabriel and Mwangwa, Florence and Ssemmondo, Emmanuel and Kironde, Joel and Kabami, Jane and Owaraganise, Asiphas and Kakande, Elijah and Ssekaynzi, Bob and Abbott, Rachel and Ayieko, James and Ruel, Theodore and Kwariisima, Dalsone and Kamya, Moses and Petersen, Maya and Havlir, Diane V and Balzer, Laura B and {the SEARCH collaboration}},
	month = jun,
	year = {2024},
	pages = {1601--1607},
}

@article{benitez_defining_2023,
	title = {Defining and estimating effects in cluster randomized trials: {A} methods comparison},
	volume = {42},
	copyright = {© 2023 The Authors. Statistics in Medicine published by John Wiley \& Sons Ltd.},
	issn = {1097-0258},
	shorttitle = {Defining and estimating effects in cluster randomized trials},
	url = {https://onlinelibrary.wiley.com/doi/abs/10.1002/sim.9813},
	doi = {10.1002/sim.9813},
	language = {en},
	number = {19},
	urldate = {2024-07-11},
	journal = {Statistics in Medicine},
	author = {Benitez, Alejandra and Petersen, Maya L. and van der Laan, Mark J. and Santos, Nicole and Butrick, Elizabeth and Walker, Dilys and Ghosh, Rakesh and Otieno, Phelgona and Waiswa, Peter and Balzer, Laura B.},
	year = {2023},
	note = {\_eprint: https://onlinelibrary.wiley.com/doi/pdf/10.1002/sim.9813},
	pages = {3443--3466},
}

@inproceedings{pearl_direct_2001,
	address = {San Francisco, CA, USA},
	series = {{UAI}'01},
	title = {Direct and indirect effects},
	isbn = {978-1-55860-800-9},
	urldate = {2024-07-24},
	booktitle = {Proceedings of the {Seventeenth} conference on {Uncertainty} in artificial intelligence},
	publisher = {Morgan Kaufmann Publishers Inc.},
	author = {Pearl, Judea},
	month = aug,
	year = {2001},
	pages = {411--420},
}

@article{martinussen_estimation_2023,
	title = {Estimation of separable direct and indirect effects in continuous time},
	volume = {79},
	issn = {1541-0420},
	doi = {10.1111/biom.13559},
	language = {eng},
	number = {1},
	journal = {Biometrics},
	author = {Martinussen, Torben and Stensrud, Mats Julius},
	month = mar,
	year = {2023},
	pmid = {34506039},
	pages = {127--139},
}

@article{stensrud_generalized_2021,
	title = {A generalized theory of separable effects in competing event settings},
	volume = {27},
	issn = {1380-7870},
	url = {https://www.ncbi.nlm.nih.gov/pmc/articles/PMC8536652/},
	doi = {10.1007/s10985-021-09530-8},
	number = {4},
	urldate = {2024-12-31},
	journal = {Lifetime Data Analysis},
	author = {Stensrud, Mats J. and Hernán, Miguel A. and Tchetgen Tchetgen, Eric J and Robins, James M. and Didelez, Vanessa and Young, Jessica G.},
	year = {2021},
	pmid = {34468923},
	pmcid = {PMC8536652},
	pages = {588--631},
}

@misc{unaids_2024_2024,
	title = {{The} {Urgency} of {Now}: {AIDS} at a {Crossroads} —  global {AIDS} report },
    year   = {2024},
	shorttitle = {2024 global {AIDS} report — {The} {Urgency} of {Now}},
	url = {https://www.unaids.org/en/resources/documents/2024/global-aids-update-2024},
    author = {UNAIDS},
	language = {en},
	urldate = {2025-01-23},
}

@article{robins_identifiability_1992,
	title = {Identifiability and exchangeability for direct and indirect effects},
	volume = {3},
	issn = {1044-3983},
	doi = {10.1097/00001648-199203000-00013},
	language = {eng},
	number = {2},
	journal = {Epidemiology (Cambridge, Mass.)},
	author = {Robins, J. M. and Greenland, S.},
	month = mar,
	year = {1992},
	pmid = {1576220},
	pages = {143--155},
}

@article{mbonye_alcohol_2014,
	title = {Alcohol consumption and high risk sexual behaviour among female sex workers in {Uganda}},
	volume = {13},
	issn = {1727-9445},
	doi = {10.2989/16085906.2014.927779},
	language = {eng},
	number = {2},
	journal = {African journal of AIDS research: AJAR},
	author = {Mbonye, Martin and Rutakumwa, Rwamahe and Weiss, Helen and Seeley, Janet},
	year = {2014},
	pmid = {25174631},
	pages = {145--151},
}

@article{kiwanuka_population_2017,
	title = {Population attributable fraction of incident {HIV} infections associated with alcohol consumption in fishing communities around {Lake} {Victoria}, {Uganda}},
	volume = {12},
	issn = {1932-6203},
	url = {https://journals.plos.org/plosone/article?id=10.1371/journal.pone.0171200},
	doi = {10.1371/journal.pone.0171200},
	language = {en},
	number = {2},
	urldate = {2025-01-23},
	journal = {PLOS ONE},
	author = {Kiwanuka, Noah and Ssetaala, Ali and Ssekandi, Ismail and Nalutaaya, Annet and Kitandwe, Paul Kato and Ssempiira, Julius and Bagaya, Bernard Ssentalo and Balyegisawa, Apolo and Kaleebu, Pontiano and Hahn, Judith and Lindan, Christina and Sewankambo, Nelson Kaulukusi},
	month = feb,
	year = {2017},
	note = {Publisher: Public Library of Science},
	pages = {e0171200},
}

@article{nyabuti_characteristics_2021,
	title = {Characteristics of {HIV} seroconverters in the setting of universal test and treat: {Results} from the {SEARCH} trial in rural {Uganda} and {Kenya}},
	volume = {16},
	issn = {1932-6203},
	shorttitle = {Characteristics of {HIV} seroconverters in the setting of universal test and treat},
	url = {https://www.ncbi.nlm.nih.gov/pmc/articles/PMC7864429/},
	doi = {10.1371/journal.pone.0243167},
	number = {2},
	urldate = {2025-01-23},
	journal = {PLoS ONE},
	author = {Nyabuti, Marilyn N. and Petersen, Maya L. and Bukusi, Elizabeth A. and Kamya, Moses R. and Mwangwa, Florence and Kabami, Jane and Sang, Norton and Charlebois, Edwin D. and Balzer, Laura B. and Schwab, Joshua D. and Camlin, Carol S. and Black, Douglas and Clark, Tamara D. and Chamie, Gabriel and Havlir, Diane V. and Ayieko, James},
	month = feb,
	year = {2021},
	pmid = {33544717},
	pmcid = {PMC7864429},
	pages = {e0243167},
}

@article{velloza_hiv-risk_2017,
	title = {{HIV}-risk behaviors and social support among men and women attending alcohol-serving venues in {South} {Africa}: {Implications} for {HIV} prevention},
	volume = {21},
	issn = {1090-7165},
	shorttitle = {{HIV}-risk behaviors and social support among men and women attending alcohol-serving venues in {South} {Africa}},
	url = {https://www.ncbi.nlm.nih.gov/pmc/articles/PMC5844773/},
	doi = {10.1007/s10461-017-1853-z},
	number = {Suppl 2},
	urldate = {2025-01-23},
	journal = {AIDS and behavior},
	author = {Velloza, Jennifer and Watt, Melissa H. and Abler, Laurie and Skinner, Donald and Kalichman, Seth C. and Dennis, Alexis C. and Sikkema, Kathleen J.},
	month = nov,
	year = {2017},
	pmid = {28710711},
	pmcid = {PMC5844773},
	pages = {144--154},
}

@article{cain_hiv_2012,
	title = {{HIV} risks associated with patronizing alcohol serving establishments in {South} {African} {Townships}, {Cape} {Town}},
	volume = {13},
	issn = {1573-6695},
	doi = {10.1007/s11121-012-0290-5},
	language = {eng},
	number = {6},
	journal = {Prevention Science: The Official Journal of the Society for Prevention Research},
	author = {Cain, Demetria and Pare, Valerie and Kalichman, Seth C. and Harel, Ofer and Mthembu, Jacqueline and Carey, Michael P. and Carey, Kate B. and Mehlomakulu, Vuyelwa and Simbayi, Leickness C. and Mwaba, Kelvin},
	month = dec,
	year = {2012},
	pmid = {22992872},
	pmcid = {PMC4540371},
	pages = {627--634},
}

@phdthesis{gupta_mechanism_2024,
	title = {Mechanism and {Mediation}: {Counterfactual} {Strata} {Effects} for {Perinatal} {Epidemiology}},
	shorttitle = {Mechanism and {Mediation}},
	url = {https://escholarship.org/uc/item/5m34j2p6},
	language = {en},
	urldate = {2025-01-24},
	school = {UC Berkeley},
	author = {Gupta, Shalika},
	year = {2024},
}

@article{balzer_evaluation_2017,
	title = {Evaluation of {Progress} {Towards} the {UNAIDS} 90-90-90 {HIV} {Care} {Cascade}: {A} {Description} of {Statistical} {Methods} {Used} in an {Interim} {Analysis} of the {Intervention} {Communities} in the {SEARCH} {Study}},
	shorttitle = {Evaluation of {Progress} {Towards} the {UNAIDS} 90-90-90 {HIV} {Care} {Cascade}},
	url = {https://biostats.bepress.com/ucbbiostat/paper357},
	journal = {U.C. Berkeley Division of Biostatistics Working Paper Series},
	author = {Balzer, Laura and Schwab, Joshua and Laan, Mark van der and Petersen, Maya},
	month = feb,
	year = {2017},
}

@article{robins_new_1986,
	title = {A new approach to causal inference in mortality studies with a sustained exposure period—application to control of the healthy worker survivor effect},
	volume = {7},
	issn = {0270-0255},
	url = {https://www.sciencedirect.com/science/article/pii/0270025586900886},
	doi = {10.1016/0270-0255(86)90088-6},
	number = {9},
	urldate = {2025-01-24},
	journal = {Mathematical Modelling},
	author = {Robins, James},
	month = jan,
	year = {1986},
	pages = {1393--1512},
}

@article{rosenblum_simple_2010,
	title = {Simple, {Efficient} {Estimators} of {Treatment} {Effects} in {Randomized} {Trials} {Using} {Generalized} {Linear} {Models} to {Leverage} {Baseline} {Variables}},
	volume = {6},
	issn = {1557-4679},
	url = {https://www.ncbi.nlm.nih.gov/pmc/articles/PMC2898625/},
	doi = {10.2202/1557-4679.1138},
	number = {1},
	urldate = {2025-01-24},
	journal = {The International Journal of Biostatistics},
	author = {Rosenblum, Michael and van der Laan, Mark J.},
	month = apr,
	year = {2010},
	pmid = {20628636},
	pmcid = {PMC2898625},
	pages = {13},
}

@book{hayes_cluster_2009,
	address = {New York},
	title = {Cluster {Randomised} {Trials}},
	isbn = {978-0-429-14205-5},
	publisher = {Chapman and Hall/CRC},
	author = {Hayes, Richard J. and Moulton, Lawrence H.},
	month = jan,
	year = {2009},
	doi = {10.1201/9781584888178},
}

@article{kakande_community-based_2023,
	title = {A community-based dynamic choice model for {HIV} prevention improves {PrEP} and {PEP} coverage in rural {Uganda} and {Kenya}: a cluster randomized trial},
	volume = {26},
	issn = {1758-2652},
	shorttitle = {A community-based dynamic choice model for {HIV} prevention improves {PrEP} and {PEP} coverage in rural {Uganda} and {Kenya}},
	doi = {10.1002/jia2.26195},
	language = {eng},
	number = {12},
	journal = {Journal of the International AIDS Society},
	author = {Kakande, Elijah R. and Ayieko, James and Sunday, Helen and Biira, Edith and Nyabuti, Marilyn and Agengo, George and Kabami, Jane and Aoko, Colette and Atuhaire, Hellen N. and Sang, Norton and Owaranganise, Asiphas and Litunya, Janice and Mugoma, Erick W. and Chamie, Gabriel and Peng, James and Schrom, John and Bacon, Melanie C. and Kamya, Moses R. and Havlir, Diane V. and Petersen, Maya L. and Balzer, Laura B. and {SEARCH Study Team}},
	month = dec,
	year = {2023},
	pmid = {38054535},
	pmcid = {PMC10698808},
	pages = {e26195},
}

@article{shahmanesh_effect_2021,
	title = {Effect of peer-distributed {HIV} self-test kits on demand for biomedical {HIV} prevention in rural {KwaZulu}-{Natal}, {South} {Africa}: a three-armed cluster-randomised trial comparing social networks versus direct delivery},
	volume = {6},
	issn = {2059-7908},
	shorttitle = {Effect of peer-distributed {HIV} self-test kits on demand for biomedical {HIV} prevention in rural {KwaZulu}-{Natal}, {South} {Africa}},
	doi = {10.1136/bmjgh-2020-004574},
	language = {eng},
	number = {Suppl 4},
	journal = {BMJ global health},
	author = {Shahmanesh, Maryam and Mthiyane, T. Nondumiso and Herbsst, Carina and Neuman, Melissa and Adeagbo, Oluwafemi and Mee, Paul and Chimbindi, Natsayi and Smit, Theresa and Okesola, Nonhlanhla and Harling, Guy and McGrath, Nuala and Sherr, Lorraine and Seeley, Janet and Subedar, Hasina and Johnson, Cheryl and Hatzold, Karin and Terris-Prestholt, Fern and Cowan, Frances M. and Corbett, Elizabeth Lucy},
	month = jul,
	year = {2021},
	pmid = {34315730},
	pmcid = {PMC8317107},
	pages = {e004574},
}

@article{ortblad_stand-alone_2023,
	title = {Stand-alone model for delivery of oral {HIV} pre-exposure prophylaxis in {Kenya}: a single-arm, prospective pilot evaluation},
	volume = {26},
	issn = {1758-2652},
	shorttitle = {Stand-alone model for delivery of oral {HIV} pre-exposure prophylaxis in {Kenya}},
	doi = {10.1002/jia2.26131},
	language = {eng},
	number = {6},
	journal = {Journal of the International AIDS Society},
	author = {Ortblad, Katrina F. and Mogere, Peter and Omollo, Victor and Kuo, Alexandra P. and Asewe, Magdaline and Gakuo, Stephen and Roche, Stephanie and Mugambi, Mary and Mugambi, Melissa Latigo and Stergachis, Andy and Odoyo, Josephine and Bukusi, Elizabeth A. and Ngure, Kenneth and Baeten, Jared M.},
	month = jun,
	year = {2023},
	pmid = {37306128},
	pmcid = {PMC10258863},
	pages = {e26131},
}

@misc{kabami_multi-component_2024,
	address = {Rochester, NY},
	type = {{SSRN} {Scholarly} {Paper}},
	title = {A {Multi}-{Component} {Integrated} {HIV} and {Hypertension} {Care} {Model} {Improves} {Hypertension} {Screening} and {Control} in {Rural} {Uganda}: {A} {Cluster} {Randomized} {Trial}},
	shorttitle = {A {Multi}-{Component} {Integrated} {HIV} and {Hypertension} {Care} {Model} {Improves} {Hypertension} {Screening} and {Control} in {Rural} {Uganda}},
	url = {https://papers.ssrn.com/abstract=5050332},
	doi = {10.2139/ssrn.5050332},
	language = {en},
	urldate = {2025-05-30},
	author = {Kabami, Jane and Balzer, Laura B. and Atukunda, Mucunguzi and Arinaitwe, Elizabeth and Mutungi, Gerald and Twinamatsiko, Brian and Mwesigye, Ronald Aine and Ayebare, Michael and Asiimwe, Alan and Akatukwasa, Cecilia and Nangendo, Joan and Shade, Starley B. and Charlebois, Edwin D. and Okello, Emmy and Kapiga, Saidi and Grosskurth, Heiner and Kamya, Moses},
	month = dec,
	year = {2024},
}

@book{fisher_statistical_1932,
  author       = {Fisher, R. A.},
  title        = {Statistical Methods for Research Workers},
  edition      = {4th, revised and enlarged},
  publisher    = {Oliver and Boyd},
  address      = {Edinburgh},
  year         = {1932},
  pages        = {xiv + 307},
  note         = {Biological Monographs and Manuals},
  url          = {https://archive.org/details/statisticalmetho00fish},
  language     = {English}
}

@article{gail_design_1996,

title   = {Design Considerations for Studies of Intervention Effects on Recurrence},
author = {Gail, Mitchell H. and Mark, Steven D. and Carroll, Raymond J. and Green, Sylvan B. and Pee, David},
  journal = {Statistics in Medicine},
  year    = {1996},
  volume  = {15},
  pages   = {123--135}
}

@article{tsiatis_covariate_2008,
  title     = {Covariate Adjustment for Two-Sample Treatment Comparisons in Randomized Clinical Trials: A Principled Yet Flexible Approach},
  author    = {Tsiatis, Anastasios A. and Davidian, Marie and Zhang, Min and Lu, Xiaomin},
  journal   = {Statistics in Medicine},
  volume    = {27},
  number    = {23},
  pages     = {4658--4677},
  year      = {2008},
  month     = {Oct},
  doi       = {10.1002/sim.3113},
  issn      = {0277-6715},
  pmid      = {17960577},
  pmcid     = {PMC2562926},
  language  = {English},
  url       = {https://www.ncbi.nlm.nih.gov/pmc/articles/PMC2562926/}
}

@article{goma_predicting_2024,
	title = {Predicting harmful alcohol use prevalence in {Sub}-{Saharan} {Africa} between 2015 and 2019: {Evidence} from population-based {HIV} impact assessment},
	volume = {19},
	issn = {1932-6203},
	shorttitle = {Predicting harmful alcohol use prevalence in {Sub}-{Saharan} {Africa} between 2015 and 2019},
	doi = {10.1371/journal.pone.0301735},
	language = {eng},
	number = {10},
	journal = {PloS One},
	author = {Goma, Mtumbi and Ng'ambi, Wingston Felix and Zyambo, Cosmas},
	year = {2024},
	pmid = {39383142},
	pmcid = {PMC11463767},
	pages = {e0301735},
}

@article{rubin_empirical_2008,
	title = {Empirical efficiency maximization: improved locally efficient covariate adjustment in randomized experiments and survival analysis},
	volume = {4},
	issn = {1557-4679},
	shorttitle = {Empirical efficiency maximization},
	language = {eng},
	number = {1},
	journal = {The International Journal of Biostatistics},
	author = {Rubin, Daniel B. and van der Laan, Mark J.},
	year = {2008},
	pmid = {19381345},
	pmcid = {PMC2669310},
	pages = {Article 5},
}

@article{li_model-robust_2025,
	title = {Model-{Robust} {Standardization} in {Cluster}-{Randomized} {Trials}},
	volume = {44},
	copyright = {© 2025 John Wiley \& Sons Ltd.},
	issn = {1097-0258},
	url = {https://onlinelibrary.wiley.com/doi/abs/10.1002/sim.70270},
	doi = {10.1002/sim.70270},
	language = {en},
	number = {20-22},
	urldate = {2025-11-08},
	journal = {Statistics in Medicine},
	author = {Li, Fan and Tong, Jiaqi and Fang, Xi and Cheng, Chao and Kahan, Brennan C. and Wang, Bingkai},
	year = {2025},
	note = {\_eprint: https://onlinelibrary.wiley.com/doi/pdf/10.1002/sim.70270},
	pages = {e70270},
}

@misc{schwab_ltmle_2023,
	title = {ltmle: {Longitudinal} {Targeted} {Maximum} {Likelihood} {Estimation}},
	copyright = {GPL-2},
	shorttitle = {ltmle},
	url = {https://cran.r-project.org/web/packages/ltmle/index.html},
	urldate = {2025-11-13},
	author = {Schwab, Joshua and Lendle, Samuel and Petersen, Maya and Laan, Mark van der and Gruber, Susan},
	month = apr,
	year = {2023},
}

@book{vanderVaart1998,
  title     = {Asymptotic Statistics},
  author    = {{van der Vaart}, Aad W.},
  year      = {1998},
  publisher = {Cambridge University Press},
  address   = {Cambridge}
}

@book{Tsiatis2006,
  title     = {Semiparametric Theory and Missing Data},
  author    = {Tsiatis, Anastasios A.},
  year      = {2006},
  publisher = {Springer},
  address   = {New York}
}

@article{schnitzer_effect_2014,
	title = {{EFFECT} {OF} {BREASTFEEDING} {ON} {GASTROINTESTINAL} {INFECTION} {IN} {INFANTS}: {A} {TARGETED} {MAXIMUM} {LIKELIHOOD} {APPROACH} {FOR} {CLUSTERED} {LONGITUDINAL} {DATA}},
	volume = {8},
	issn = {1932-6157},
	shorttitle = {{EFFECT} {OF} {BREASTFEEDING} {ON} {GASTROINTESTINAL} {INFECTION} {IN} {INFANTS}},
	doi = {10.1214/14-aoas727},
	language = {eng},
	number = {2},
	journal = {The Annals of Applied Statistics},
	author = {Schnitzer, Mireille E. and van der Laan, Mark J. and Moodie, Erica E. M. and Platt, Robert W.},
	month = jun,
	year = {2014},
	pmid = {25505499},
	pmcid = {PMC4259272},
	pages = {703--725},
}

@misc{unaids_2025_2025,
  title        = {{AIDS}, crisis and the power to transform: {UNAIDS} Global {AIDS} Update},
  year         = {2025},
  organization = {{Joint United Nations Programme on HIV/AIDS}},
  address      = {Geneva},
  url          = {https://www.unaids.org/en/resources/documents/2025/2025-global-aids-update},
  author = {UNAIDS},
  note         = {Licence: CC BY-NC-SA 3.0 IGO},
  language     = {en},
  urldate      = {2025-01-23}
}

@misc{unaids_2021_prevention_cascades,
  title        = {Creating HIV prevention cascades: Operational guidance on a tool for monitoring programmes},
  year         = {2021},
  organization = {{Joint United Nations Programme on HIV/AIDS (UNAIDS)}},
  address      = {Geneva},
  url          = {https://www.unaids.org/sites/default/files/media_asset/JC3038_creating-hiv-prevention-cascades_en.pdf},
  note         = {Accessed 2025-01-23},
  author = {UNAIDS},
  language     = {en}
}

@techreport{unaids_hiv_nodate,
	title = {The {HIV} test-and-treat cascade – {AIDS} },
      year         = {2020},
	url = {https://aids2020.unaids.org/chapter/chapter-2-2020-commitments/the-hiv-test-and-treat-cascade/},
	language = {en},
	urldate = {2026-02-19},
	author = {UNAIDS},
}

@misc{beesiga_IAS_2025,
  author = {Beesiga, Brian and Litunya, Janice and Nakato, Joy Z. and Agola, Jaquiline A. and Marson, Kara and Mugoma, Wafula E. and Agaba, Pius and Temple, Jennifer and Camlin, Carol S. and Shade, Starley B. and Woolf-King, Sarah E. and Hahn, Judith A. and Kakande, Elijah and Kabami, Jane and Petersen, Maya L. and Havlir, Diane V. and Balzer, Laura B. and Ayieko, James and Kamya, Moses R. and Chamie, Gabriel and {OPAL-East Africa trial team}},
  title  = {Higher levels of alcohol use associated with increased HIV risk among persons at alcohol-serving venues in the OPAL trial in Kenya and Uganda},
  year   = {2025},
  note   = {Poster presented at the International AIDS Conference, Kigali, Rwanda}
}

@misc{Litunya_IAS_2024,
  author = {Litunya, Janice and Beesiga, Brian and Nakato, Joy Z. and Agola, Jaquiline A. and Marson, Kara and Mugoma, Wafula E. and Temple, Jennifer and Camlin, Carol S. and Shade, Starley B. and Hahn, Judith A. and Kakande, Elijah and Kabami, Jane and Petersen, Maya L. and Havlir, Diane V. and Kamya, Moses R. and Balzer, Laura B. and Ayieko, James and Chamie, Gabriel},
  title  = {Impact of alcohol drinking venue characteristics on yield of HIV status-neutral testing in rural East Africa},
  year   = {2024},
  note   = {Poster presented at the International AIDS Conference, Munich, Germany}
}

@article{fiero_statistical_2016,
	title = {Statistical analysis and handling of missing data in cluster randomized trials: a systematic review},
	volume = {17},
	issn = {1745-6215},
	shorttitle = {Statistical analysis and handling of missing data in cluster randomized trials},
	url = {https://pmc.ncbi.nlm.nih.gov/articles/PMC4748550/},
	doi = {10.1186/s13063-016-1201-z},
	urldate = {2026-02-19},
	journal = {Trials},
	author = {Fiero, Mallorie H. and Huang, Shuang and Oren, Eyal and Bell, Melanie L.},
	month = feb,
	year = {2016},
	pmid = {26862034},
	pmcid = {PMC4748550},
	pages = {72},
}

@article{liang_longitudinal_1986,
	title = {Longitudinal data analysis using generalized linear models},
	volume = {73},
	issn = {0006-3444},
	url = {https://doi.org/10.1093/biomet/73.1.13},
	doi = {10.1093/biomet/73.1.13},
	number = {1},
	urldate = {2026-02-19},
	journal = {Biometrika},
	author = {LIANG, KUNG-YEE and ZEGER, SCOTT L.},
	month = apr,
	year = {1986},
	pages = {13--22},
}

@misc{balzer_SER_2025,
  author       = {Joy Z Nakato and Laura B. Balzer },
  title        = {Adaptive Pooling to minimize bias and maximize power in cluster-randomized trials },
  year         = {2025},
  note         = {Presentation at the Soceity for Epidemiological Research (SER) Conference  2025, Boston, MA}
}

@article{hickey_community_2025,
	title = {Community health worker–facilitated telehealth for moderate–severe hypertension care in {Kenya} and {Uganda}: {A} randomized controlled trial},
	volume = {22},
	issn = {1549-1676},
	shorttitle = {Community health worker–facilitated telehealth for moderate–severe hypertension care in {Kenya} and {Uganda}},
	url = {https://journals.plos.org/plosmedicine/article?id=10.1371/journal.pmed.1004632},
	doi = {10.1371/journal.pmed.1004632},
	language = {en},
	number = {6},
	urldate = {2026-02-27},
	journal = {PLOS Medicine},
	author = {Hickey, Matthew D. and Owaraganise, Asiphas and Ogachi, Sabina and Sang, Norton and Wafula, Erick M. and Kabami, Jane and Sutter, Nicole and Temple, Jennifer and Muiru, Anthony and Chamie, Gabriel and Kakande, Elijah and Petersen, Maya L. and Balzer, Laura B. and Havlir, Diane V. and Kamya, Moses R. and Ayieko, James},
	month = jun,
	year = {2025},
	note = {Publisher: Public Library of Science},
	pages = {e1004632},
}

\clearpage

\end{document}